\documentclass[%
superscriptaddress,
reprint,
 amsmath,amssymb,
aps,
prl,
floatfix]{revtex4-2}

\usepackage{bbold}
\usepackage{graphicx}
\usepackage{dcolumn}
\usepackage{bm}
\usepackage{setspace}
\usepackage{mathrsfs}
\usepackage{hyperref}

\usepackage{graphicx}
\usepackage{dcolumn}
\usepackage{bm}
\usepackage[normalem]{ulem}

\usepackage{amsmath}
\usepackage[caption=false]{subfig}

\usepackage{braket}

\usepackage{pgfplots}
\pgfplotsset{width=8.6cm,compat=1.5}

\usepackage{textcomp}

\usepackage{xcolor}

\usepackage{hyperref}
\hypersetup{
    colorlinks=true,
    linkcolor=blue,
    citecolor=blue,      
    urlcolor=blue,
}

\usepackage{csquotes}

\usepackage{bbold}

\usepackage{mathrsfs}

\usepackage[rightcaption]{sidecap}

\graphicspath{{Pictures/}}

\begin{document}

\preprint{APS/123-QED}

\title{Entanglement witness for combined atom interferometer-mechanical oscillator setup}

\author{Gayathrini Premawardhana}
\email{gtp6626@umd.edu}
\affiliation{Joint Center for Quantum Information and Computer Science, University of Maryland-NIST, College Park, Maryland 20742, USA}
\author{Deven P. Bowman}
\email{dbow@stanford.edu}
\affiliation{Department of Physics, University of Maryland, College Park, Maryland 20742, USA}
\affiliation{Department of Physics, Stanford University, Stanford, California 94305, USA}
\author{Jacob M. Taylor}
\email{jmtaylor@umd.edu}
\affiliation{Joint Center for Quantum Information and Computer Science, University of Maryland-NIST, College Park, Maryland 20742, USA}
\affiliation{Joint Quantum Institute, University of Maryland-NIST, College Park, Maryland 20742, USA}





\date{\today}


\begin{abstract}

We investigate how to entangle an atom interferometer and a macroscopic mechanical oscillator in order to create non-classical states of the oscillator. We propose an entanglement witness, from whose violation, the generation of entanglement can be determined. We do this for both the noiseless case and when including thermal noise. Thermal noise can arise from two sources: the first being when the oscillator starts in a thermal state, and the second when a continuous thermal bath is in contact with the oscillator. We find that for the appropriate oscillator quality factor $Q$, violation always exists for any value of magnetic coupling $\lambda$ and thermal occupancy $\Bar{n}$. Cooling the oscillator to its ground state provides an $O(\bar n)$ improvement in the EW violation than starting in a thermal state. However, this still requires at least 10$^{5}$ measurements to be determined. We then consider how to use magnetic interactions to realize this idea.

\end{abstract}

\maketitle


Entanglement is a core feature of quantum mechanics that precludes any hidden variable classical theory. Many groups have considered entanglement in macroscopic systems, particularly in the case of spin-spin systems \cite{KongVaporEnt,BoseEW,SpinSpin2,SpinSpin3,SpinSpin4}, oscillator-oscillator systems \cite{OckeloenMassiveMechOsc, RiedingerRemoteEnt,OscOsc1,OscOsc2,OscOsc3,OscOsc4,OscOsc5}, and spin-oscillator systems \cite{ThomasMechSpinEnt,SpinOsc1,SpinOsc2}. For instance, Kong et al. experimentally generate entanglement between the spins of atoms in a vapor \cite{KongVaporEnt}, Ockeloen-Korppi et al. use a cavity to entangle two mirrors located 600 µm apart \cite{OckeloenMassiveMechOsc}, and Riedinger et al. entangle two chips separated by 20 cm \cite{RiedingerRemoteEnt}. Thomas et al.  entangle a spin ensemble with a square membrane of sides 1 mm in length \cite{ThomasMechSpinEnt}. 

Of particular interest are approaches to entangle masses and/or spins to investigate theories that incorporate quantum mechanics with gravity. For example, Kafri et al. \cite{kafri2013noiseinequalityclassicalforces}, Bose et al. \cite{BoseEW} and Marletto et al. \cite{MarlettoGravEnt} propose using two masses which interact with each other though a gravitational force; if the two masses become entangled, then we can conclude that the gravitational field is capable of generating entanglement. A different approach by Carney et al. \cite{PRXQuantum.2.030330} suggests gravitationally coupling an atom interferometer with a mechanical oscillator. If the interaction is capable of entanglement, the oscillator and the atoms will get entangled, which can be detected using the interferometer's signal.

Such experiments have to use certain techniques to determine the existence of entanglement. The experiments in Ref. \cite{KongVaporEnt,OckeloenMassiveMechOsc,RiedingerRemoteEnt,ThomasMechSpinEnt,BoseEW} use (or propose using) an entanglement witness (EW) that work for very low noise situations, where one prepares pure states of the two systems, and noise and decoherence are neglected. An EW is a quantity composed of observables; it has a bound satisfied by separable states, which, if violated, determines that entanglement exists between the systems. 
The literature on creating an entanglement witness is varied. The most well-known example of EWs might be Bell inequalities \cite{BellInequality,GuhneEntDetection} used to distinguish between the possibility of local hidden variable theories and entanglement. Another common EW is one which is often applied to oscillator-oscillator entanglement, whose general version was proposed by Duan et al. \cite{DuanInsepCriterion}.

In this paper, we work within the context of the proposal by Carney et al. \cite{PRXQuantum.2.030330}, introduced previously. In the progress towards such a gravity experiment, it is more reasonable to first attempt the generation of \textit{magnetic} entanglement between the atom interferometer and the oscillator, since a sufficiently strong magnetic coupling is easier to obtain. Before proceeding with physically building the experiment, it is important to ensure that entanglement will exist, at least theoretically, and understand the physical parameters at which it will occur. In order to do this, we construct an entanglement witness (EW) for our setup that provides a path for violation at finite motional temperatures and with noise and dephasing. We investigate the effects of thermal noise on its violation and establish the conditions on parameters such that violation exists. We find that for appropriate values of oscillator quality factor $Q$, violation always exists for any value of magnetic coupling $\lambda$ and thermal occupancy $\Bar{n}$. 

\begin{figure}[htb]
         \includegraphics[width = 8.6 cm]{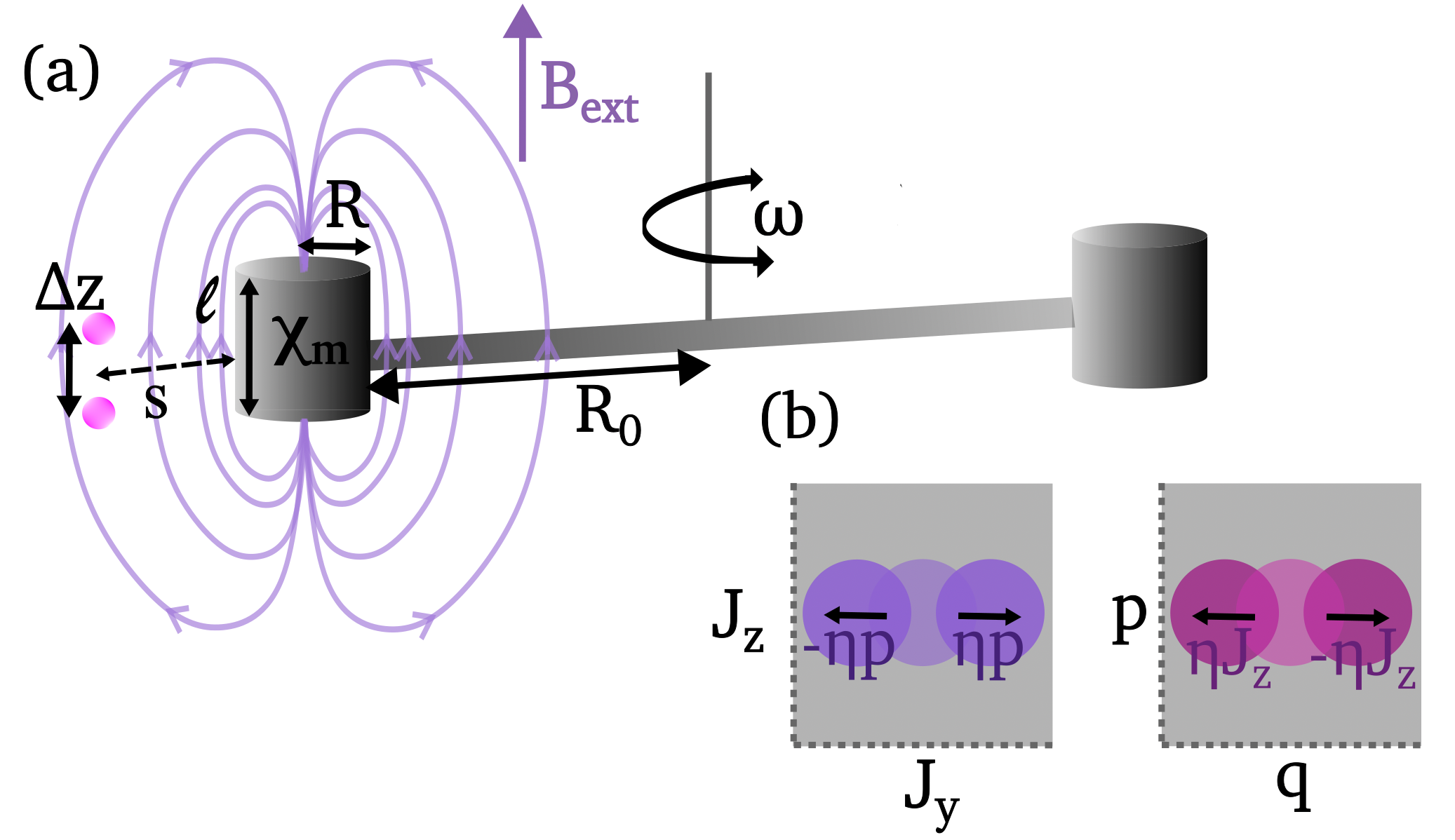}
       \caption{(a) Simplified version of oscillator-interferometer setup. The atoms (pink) of the interferometer are in traps \cite{AtomInterExpHolger}, and are located at equal distances above and below the center of the cylinder. An external magnetic field $\mathbf{B_{\rm{ext}}}$ induces a magnetic field in the cylinders which have magnetic susceptibility $\chi_{m}$. The oscillator rotates in the horizontal plane with frequency $\omega$. (b) Depiction of how the oscillator and spin operators become correlated with each other, with $\eta = \sqrt{2/(m\omega)}~(g/\omega)$. The explicit expressions are given in the Supplemental Material.}
       \label{fig:OscillatorGeoPRL}
\end{figure}

\textbf{\textit{General entanglement witness - }}
A simplified setup that we can consider is shown in Figure \ref{fig:OscillatorGeoPRL}. The Hamiltonian of the system is (note that $\hbar = 1$ herein)  \cite{PRXQuantum.2.030330}
\begin{equation}
    H =  \omega \hat c^\dagger \hat c + g (\hat c + \hat c^\dagger)\hat J_{z}. \label{eq:HamiltMultiSpin}
\end{equation}
$\hat c^\dagger$ and $\hat c$ are the oscillator creation and annihilation operators. $\hat J_{z}$ is a pseudo-spin operator describing the location of $N$ atoms; that is, $\hat{J}_{z}=\sum_{j=1}^{N} \hat S_{z,j}$, where $\hat S_{z,j}$ characterizes which of the two interferometer traps the $j^{th}$ atom is in. As stated earlier, the creation of entanglement between the interferometer and the oscillator can be determined by measuring an entanglement witness (EW).
We find that the following quantity is an applicable EW for our setup:
\begin{equation}
    W = \textrm{Var}(\hat{J}_{x})+\textrm{Var}(\hat{J}_{y}+a_{y}\hat{q}+b_{y}\hat{p})+\textrm{Var}(\hat{J}_{z}+a_{z}\hat{q}+b_{z}\hat{p})
    \label{eq:EWmain}
\end{equation}
Here, $\hat{J}_{\mu}=\sum_{j=1}^{N} \hat S_{\mu,j}$ and $\hat q(t) = \sqrt{1/(2 m \omega)}(\hat c + \hat c^{\dagger})$. The coefficients $a_{\mu}$ and $b_{\mu}$ can be chosen as appropriate. Such spin-oscillator variances allow for the characterization of spin-oscillator correlations.
Eq. \ref{eq:EWmain} is bounded by $W_{b}$ for all separable states. If this bound is violated by the experimental state of concern, we can conclude that it is an entangled state.

$W_{b}$ can be derived using the theorem in the paper by Hofmann et al.~\cite{EWboundMain}. They showed that for operators $\hat{A}_{\mu}$ of system $A$ and $\hat{B}_{\mu}$ of system $B$, if $\sum_{\mu}\textrm{Var}(\hat{A}_{\mu}) \geq U_{A}$ and $\sum_{\mu}\textrm{Var}(\hat{B}_{\mu}) \geq U_{B}$, then $\sum_{\mu}\rm{Var}(\hat{A}_{\mu}+\hat{B}_{\mu}) \geq U_{A}+U_{B}$ for any separable state. Starting with $\tilde{W}=\sum_{\mu}\textrm{Var}(\hat{J}_{\mu}+a_{\mu}\hat{q}+b_{\mu}\hat{p})$, where the summation $\sum_{\mu}$ is over the $x,y,$ and $z$ components, we follow Ref.~\cite{EWboundMain} to arrive at the general bound $ \tilde{W} \geq \left[\sum_{\mu}\textrm{Var}(\hat{J}_{\mu})\right]_{\rm{min}}+\left[\sum_{\mu}\textrm{Var}(a_{\mu}\hat{q}+b_{\mu}\hat{p})\right]_{\rm{min}}$. To obtain the witness in Eq. \ref{eq:EWmain}, we set $a_{x}=0$ and $b_{x}=0$ on the left and right hand sides of this inequality.

The oscillator-operator bound can be derived using the Heisenberg uncertainty principle and carrying out a minimization process; we get $\sum_{\mu \neq x}\textrm{Var}(a_{\mu}\hat{q}+b_{\mu}\hat{p}) \geq |a_{y}b_{z}-a_{z}b_{y}|$. In analogy to a derivation in Ref.~\cite{EWboundMain}, the general bound for the spin operators is $\sum_{\mu}\textrm{Var}(\hat{J}_{\mu}) \geq j$, which for a fully symmetric state will be $\sum_{\mu}\textrm{Var}(\hat{J}_{\mu}) \geq N/2$. The most general case will be $\sum_{\mu}\textrm{Var}(\hat{J}_{\mu}) \geq [j]_{\textrm{min,exp}}$
where $[j]_{\textrm{min,exp}}$ is the minimum value that $j$ can take at the time of measurement.

For the EW in Eq. \ref{eq:EWmain}, we can finally write the general bound to be
\begin{equation}
    W_{b}= [j]_{\textrm{min,exp}}+|a_{y}b_{z}-a_{z}b_{y}|.
\end{equation} 
That is, for a separable state, the condition that $W \ge W_{b}$ will be satisfied. If we evaluate $W$ with respect to a possibly entangled state in order to obtain the value $W_{\rm{en}}$ and find that $W_{\rm{en}}<W_{b}$, then the bound is violated and we can conclude that the state is indeed entangled. 

We explore this EW in the context of two scenarios; when the systems are noiseless and when there is thermal noise (the case of atomic dephasing can be found in the Supplementary Material). For each scenario, we determine whether the EW is violated. In an experimental setting, $W_{b}$ is a theoretical value that we have at hand, while $W_{\rm{en}}$ is the quantity that we estimate through measurement. We consider the quantity $W_{\rm{ratio}}=\left[(W_{b}-W_{\rm{en}})/{W_{b}}\right]_{\rm{max}}$;
the EW is violated if $W_{\rm{ratio}}>0$. Since the imprecision is set by $W_{b}-W_{\rm{en}}$, $W_{\rm{ratio}}$ informs us of the number of measurements $n_{\rm{meas}}$ required to determine the existence of entanglement, since $n_{\rm{meas}}\sim W_{\rm{ratio}}^{-2}$.

A key step in this process is determining the coefficients $a_{\mu}$ and $b_{\mu}$, where we attempt to minimize the variance of the EW.
When $|W_{b}-W_{\rm{en}}| \ll |W_{b}|$, finding $a_{\mu}$ and $b_{\mu}$ that maximizes $|W_{b}-W_{\rm{en}}|$ suffices. Practically, however, finding $|W_{b}-W_{\rm{en}}|_{\rm{max}}$ is complicated due to the absolute value term given by $|a_{y}b_{z}-a_{z}b_{y}|$. Therefore, in this paper we minimize the value $W_{\rm{en}}$.

\textbf{\textit{Noiseless entanglement witness - }}First, we investigate the violation for the noiseless scenario. Here the atomic states starts in the $\hat J_{x}$ eigenstate $\ket{++...+}$ and the oscillator in the ground state $\ket{0}$ \cite{PRXQuantum.2.030330}; after evolving under Eq. \ref{eq:HamiltMultiSpin}, the final entangled state is given by $\ket{\psi_{\rm{en}}}$ (explicit form is given in the Supplemental Material). 

We can understand the evolution of the system described by Eq. \ref{eq:HamiltMultiSpin} within the Heisenberg picture. The first component to consider is the Heisenberg operator $\hat c(t) = e^{-i\omega t} \hat c(0)-i\int_{0}^{t} dt'~ g\hat J_{z}e^{-i\omega (t-t')}$. The second component is the rotation of the spin operators $\hat{J}_{x}$ and $\hat{J}_{y}$ about the z-axis by angle $\epsilon(t) = g \int_{0}^{t}dt'~ [\hat c(t')+\hat c(t')^\dagger]$, due to the term $g ~ \hat q\hat J_{z}$ in the Hamiltonian; this is visually shown in Figure \ref{fig:OscillatorGeoPRL}. 
These two components show us that the different operators become correlated with each other, for instance, $\hat q(t)$ with $\hat J_{z}$. Optimizing the values of $a_{\mu}$ and $b_{\mu}$ translates to controlling the contributions of these correlations such that the variances of Eq. \ref{eq:EWmain} are minimized for the entangled state. For instance, if we consider the operator combination $\hat{J}_{y}(t)+a_{y}\hat{q}(t)+b_{y}\hat{p}(t)$, to first order in $\lambda$ (where $\lambda=g/\omega$) for simplicity, we have
\begin{equation}
\begin{split}
    & \hat J_{y}(t)+a_{y} \hat q(t)+b_{y} \hat p(t)\\ &\approx \hat J_{y} + \lambda\hat J_{x}\left[\sqrt{\frac{2}{m \omega}}~[\hat p- \hat p \cos(\omega t)]+\sqrt{2 m \omega}~ \hat q \sin(\omega t))\right]\\&+a_{y}\left[\hat q \cos(\omega t)+\frac{\hat p \sin(\omega t)}{m \omega}\right]
    +b_{y}[\hat p \cos(\omega t)-m \omega \hat q \sin(\omega t)]\\&+O(\lambda \hat J_{z})
    \label{eq:VarJyOp}
\end{split}
\end{equation}
The first line of Eq. \ref{eq:VarJyOp} is the Heisenberg-evolved operator $\hat J_{y}(t)$ expanded to $O(\lambda)$ (with the squeezing term dropped). The term $a_{y} \hat q(t)+b_{y}\hat{p}(t)$ added to $J_{y}(t)$ contributes the coefficients $a_{y}$ and $b_{y}$; given the explicit form of $a_{y} \hat q(t)+b_{y}\hat{p}(t)$ as shown in the second line of Eq. \ref{eq:VarJyOp}, it seems plausible that values for $a_{y}$ and $b_{y}$ can be chosen such that the contribution to $\hat{J}_{y}(t)$ from $\hat{J}_{x}$ can be canceled. Note that, since we are starting in the state $\ket{++...}$, the $\hat J_{x}$ operators here evaluate to $N/2$ and the $O(\lambda \hat J_{z})$ terms evaluate to 0.

If we proceed with solving for $a_{\mu}$ and $b_{\mu}$, we see that this does indeed occur. We first expand various products of the operators ($\hat{J}_{x}(t)  \hat q(t)$, etc.) to $O(\lambda^{2})$; $O(\lambda^{2})$ should be appropriate, since the only other parameter that can greatly ``cancel out" the effects of $\lambda$ is the number of atoms $N$ and the highest order of $N$ is $N^{2}$ (which arises from the squaring of the spin operators). Once we have these Heisenberg-evolved operators and their products, we can evaluate their expectation values with respect to $\ket{++...}\ket{0}$ (which is the remainder of $\ket{\psi_{\rm{en}}}$) to obtain $W_{\rm{en}}$.

Now we can minimize $W_{\rm{en}}$ with respect to $a_{\mu}$ and $b_{\mu}$. We find that
\begin{align}
    a_{\rm{y,0,\lambda}}&=-\lambda N \sqrt\frac{m\omega}{2}\sin(\omega t)
    \label{eq:ay}\\
    b_{\rm{y,0,\lambda}}&=\lambda N\sqrt{\frac{1}{2m\omega}}[1-\cos(\omega t)]
    \label{eq:by} \\
    a_{\rm{z,0,\lambda}}&=\lambda N \sqrt\frac{m\omega}{2}[1-\cos(\omega t)]
    \label{eq:az} \\
b_{\rm{z,0,\lambda}}&=\lambda N\sqrt{\frac{1}{2m\omega}}\sin(\omega t)
    \label{eq:bz}
\end{align}
As can be seen, these coefficients and, thereby, the EW of Eq. \ref{eq:EWmain}, depend on the following experimental parameters: the  coupling $\lambda$, the number of atoms $N$, the mass of the oscillator $m$, oscillator frequency $\omega$, and the interaction time $t$. Furthermore, these coefficients satisfy $\mathbf{a} \cdot \mathbf{b} = 0$.

We can substitute these coefficients into Eq. \ref{eq:VarJyOp} to see how they can be used to cancel out correlations between operators. If we were to set $\bra{++...}\hat J_{x} \ket{++...}= N/2$, we find that $\bra{\psi_{\rm{en}}}  \hat J_{y}+a_{y} \hat q +b_{y} \hat p \ket{\psi_{\rm{en}}} = \bra{0} \bra{++...} \hat J_{y} \ket{++...}\ket{0}$. 

\begin{SCfigure*}[0.7][h]
   \caption{The black line depicts the noiseless case. We have set $\lambda = 10^{-4.5}$, $N = 10^{6}$, and $\omega = 2\pi/20$ rad/s. White noise depends only on $\bar n/Q$. Thus, the scenario where the oscillator starts in a ground state and then remains in contact with a bath can be characterized solely using $\bar n/Q$. However, a value for $\bar n$ must be set when starting in a thermal state. Since the condition $\bar n \lambda^{2} \ll 1$ must be satisfied, we consider $\bar n = 325$ to be the highest appropriate value. Purple lines correspond to $\bar n/Q = 1/10$ and blue lines correspond to $\bar n/Q = 1/4$. As can be seen in the plots, the lower the value of $\bar n/Q$, the better the violation. Note that introducing a bath results in the violation \textit{not} being seen for the entire oscillator period; where the plots get truncated is when $W_{b}-W_{\rm{en}}<0$.}
   \includegraphics[width = 0.65\textwidth]{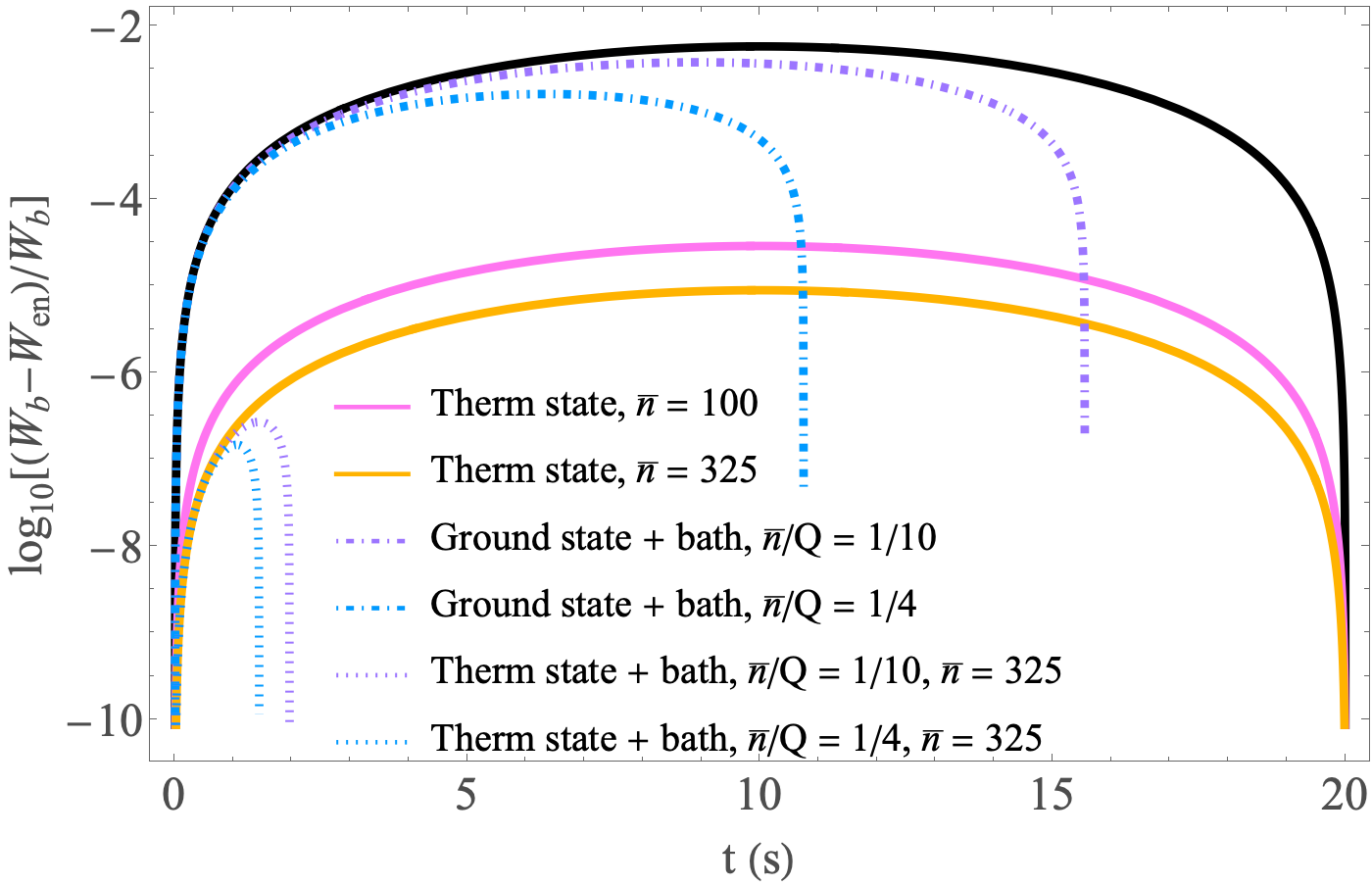}
       \label{fig:AllThermPlots}
\end{SCfigure*}

We can now finally obtain the bound
\begin{equation}
    W_{\rm{b,0}} = \frac{N}{2}+\lambda^{2}N^{2}|[1-\cos(\omega t)]|
    \label{eq:EWexpBound}
\end{equation}
and the entangled-state value 
\begin{equation}
    W_{\rm{en,0}}=\frac{N}{2}-\frac{\lambda^{2}N^{2}}{2}[1-\cos(\omega t)].
    \label{eq:EWvalueEntState}
\end{equation}
We see that $\ket{\psi}_{\rm{en}}$ violates the EW bound by a value of $(3/2)\lambda^{2}N^{2}[1-\cos(\omega t)]$. It must be emphasized again that these calculations hold for $\lambda \ll 1$. The exact limit on $\lambda$ is such that Eq. \ref{eq:EWvalueEntState} is always positive; that is, the result is valid for $\lambda <1/\sqrt{2 N}$, which for $N=10^6$ would be $\lambda < 7 \times 10^{-4}$. Fig. \ref{fig:AllThermPlots} shows a plot of the violation (black line); it is highest at $t = \pi/\omega$, which is consistent with the results in Carney et al. \cite{PRXQuantum.2.030330} where the oscillator is completely entangled with the atoms at $t = \pi/\omega$.

\textbf{\textit{Entanglement witness with a thermal state and white noise bath - }} We can now consider the case of thermal noise. There are three scenarios that can be considered; the first when the oscillator starts in a thermal state, the second when the oscillator starts in a thermal state and remains in contact with a white noise bath, and the third when the oscillator starts in the \textit{ground} state and remains in contact with a bath.

For a thermal state $\ \Hat{\rho} = \sum_{n=0}^{\infty} \frac{e^{- n \beta \omega}}{1-e^{-  \beta \omega}} \ket{n}\bra{n}$ \cite{ThermState}, we find that, to $O(\lambda)$, the new coefficients are 
\begin{align}
    a_{\rm{y,\Bar{n}}}&=-\lambda N \sqrt\frac{m\omega}{2}\sin(\omega t)
    \label{eq:ayThermState}\\
    b_{\rm{y,\Bar{n}}}&=\lambda N\sqrt{\frac{1}{2m\omega}}[1-\cos(\omega t)]
    \label{eq:byThermState} \\
    a_{\rm{z,\Bar{n}}}&=\frac{\lambda N}{2 \Bar{n} + 1} \sqrt\frac{m\omega}{2}[1-\cos(\omega t)]
    \label{eq:azThermState} \\
    b_{\rm{z,\Bar{n}}}&=\frac{\lambda N}{2 \Bar{n} + 1}\sqrt{\frac{1}{2m\omega}}\sin(\omega t)
    \label{eq:bzThermState}
\end{align}
The next order of these coefficients is $O(\lambda^{3})$. Comparing with Equations \ref{eq:ay}-\ref{eq:bz}, it can be seen that $a_{\rm{z,\Bar{n}}}$ and $b_{\rm{z,\Bar{n}}}$ have been modified by a factor of $1/(2 \Bar{n}+1)$. However, they no longer satisfy $\mathbf{a} \cdot \mathbf{b} = 0$. As explained previously, and as can be seen by their similar forms, these coefficients work together with the covariances and correlations to minimize the value of $W_{\rm{en}}$ when starting in a thermal state. 

The bound to $O(\lambda^{2})$ is 
\begin{equation}
    W_{\rm{b,\Bar{n}}} = \frac{N}{2}+\frac{1}{2 \Bar{n}+1}\lambda^{2}N^{2}|[1-\cos(\omega t)]|
    \label{eq:EWboundThermState}
\end{equation}
and the entangled state value of the EW to $O(\lambda^{2})$  is 
\begin{equation}
    W_{\rm{en,\Bar{n}}}=\frac{N}{2}-\frac{1}{2 \Bar{n}+1}\frac{\lambda^{2}N^{2}}{2}[1-\cos(\omega t)].
    \label{eq:EWvalueEntThermState}
\end{equation}
The expansion to $O(\lambda^{2})$ is only valid if $\bar n \lambda^{2} \ll 1$; larger $\bar n \lambda^{2}$ leads to key corrections associated with atomic phase evolution of order unity. Plots for different values of $\Bar{n}$ are shown in Fig. \ref{fig:AllThermPlots}; the higher the value of $\bar n$, the weaker the violation. 

Next, we consider a scenario where the oscillator is in continuous contact with white noise, which approximates a thermal bath at high temperature. The Hamiltonian is now given by
\begin{equation}
    H =  \omega \hat c^\dagger \hat c + [gJ_{z}+F_{\rm{in}}(t)](\hat c + \hat c^\dagger), 
    \label{eq:HamiltBath}
\end{equation}
where $F_{\rm{in}}(t)$ is a classical white noise force; this Hamiltonian is simply that of Eq. \ref{eq:HamiltMultiSpin} to which the effect from $F_{\rm{in}}(t)$ has been added. The 
auto-correlation function is a delta function with an amplitude set by the fluctuation-dissipation theorem $\langle\langle F_{\rm{in}}(t)F_{\rm{in}}(t')\rangle\rangle=\Bar{n}\gamma\delta(t-t')$.
Proceeding to understand within the Heisenberg picture as previously, we can solve for $\hat{c}(t)$ to obtain $\hat c(t) = e^{-i\omega t} \hat c(0)-i\int_{0}^{t} dt'~ (g\hat J_{z}+F_{in}(t'))e^{-i\omega (t-t')}$. With $\hat q(t)=q_{\rm{zpf}}[\hat c(t) + \hat c^{\dagger}(t)]$ and $\hat p(t)=i p_{\rm{zpf}}[-\hat c(t) + \hat c^{\dagger}(t)]$, the noise contributions to $ \hat q(t)$ and $\hat p(t)$ can be expressed as $\mathcal{Q}(t) = -2 q_{\rm{zpf}} \int_{0}^{t}dt'~ F_{\rm{in}}(t') \sin[\omega (t-t')]$ and $\mathcal{P}(t) = -2 p_{\rm{zpf}} \int_{0}^{t}dt'~ F_{\rm{in}}(t') \cos[\omega (t-t')]$.
Further, $\hat{J}_{x}$ and $\hat{J}_{y}$ will rotate about the z-axis by an angle $\xi(t) = g \int_{0}^{t}dt'~ [\hat c(t')+\hat c(t')^\dagger]$ as follows: $\hat{J}_{y}(t) = \hat{J}_{y} \cos[\xi(t)] + \hat{J}_{x} \sin[\xi(t)]$ and $\hat{J}_{x}(t) = \hat{J}_{x} \cos[\xi(t)] - \hat{J}_{y} \sin[\xi(t)]$. The white noise contribution to this angle is given by $\Xi(t) = -g \int_{0}^{t} dt'~ \mathcal{Q}(t')/q_{\rm{zpf}}
 = 2 g \int_{0}^{t}\int_{0}^{t'} dt''dt'~F_{\rm{in}}(t'')\sin[\omega(t'-t'')]$. Note that here, $q_{\rm{zpf}} = \sqrt{1/(2 m \omega)}$ and $p_{\rm{zpf}} = \sqrt{m \omega/2}$.

Note that $\mathcal{Q}(t)$, $\mathcal{P}(t)$, and $\Xi(t)$ are not operators since they arise from the classical noise $F_{\rm{in}}(t)$; therefore, their covariances consist of averages over classical noise and are hence represented with the notation $\langle \langle ... \rangle \rangle$. For instance, $\langle\langle \Xi(t)\Xi(t) \rangle\rangle$ contributes to the noise in the EW term $\Delta\langle\langle\hat{J}_{\rm{x}}\rangle\rangle^{2}$ and $\langle\langle \mathcal{Q}(t)\mathcal{Q}(t) \rangle\rangle$ contributes to the noise in $\langle\langle \hat q^{2} \rangle\rangle$. Therefore, the expectation values that compose the EW must be modified by adding the noise contributions  $\Delta\langle\langle\hat{J}_{\rm{x}}\rangle\rangle^{2}$, $\Delta\langle\langle q^{2}\rangle\rangle$, $\Delta\langle\langle p^{2}\rangle\rangle$, $\Delta\langle\langle \hat p \hat J_{y}+\hat J_{y} \hat p\rangle\rangle$, $\Delta\langle\langle \hat q \hat J_{y}+\hat J_{y} \hat q \rangle\rangle$ and $\Delta\langle \langle \hat q \hat p + \hat q \hat p \rangle \rangle$; an appropriate linear combination of these gives $\Delta W_{\rm{\bar n,noise}}$. Thus, we now have, $W_{\rm{en,\bar n}, noise}=W_{\rm{en,\Bar{n}}}+\Delta W_{\rm{\bar n,noise}}$, where 
\begin{equation} 
\begin{split}                  
    \Delta W_{\bar n, \rm{noise}} &= \frac{N^{2}}{4}\frac{\lambda^{2}\Bar{n}}{Q}[6 \omega t - 8 \sin(\omega t)+\sin(2 \omega t)]\\ & + a_{y} \sqrt{\frac{1}{2 m \omega}} \frac{N}{2} \frac{16 \lambda \Bar{n}}{Q}\sin^{4}\left(\frac{\omega t}{2}\right)\\
    & - b_{y} \sqrt{\frac{m \omega}{2}} \frac{N}{2} \frac{8\lambda\Bar{n}}{Q}\left[\frac{\omega t}{2}-\sin(\omega t)+\frac{\sin(2 \omega t)}{4}\right]\\ & + (a_{y}^{2}+a_{z}^{2}) \frac{1}{2 m \omega}\frac{\Bar{n}}{Q}[2 \omega t-\sin(2 \omega t)] \\
    & + (b_{y}^{2}+b_{z}^{2})\frac{m \omega}{2}\frac{\Bar{n}}{Q}[2 \omega t+\sin(2 \omega t)]\\ & + (a_{y} b_{y}+a_{z} b_{z})\frac{2 \Bar{n}}{Q}\sin^{2}(\omega t)
    \label{eq:EWbathNoise}
\end{split}  
\end{equation} 
with the coefficients being those in Equations \ref{eq:ayThermState}-\ref{eq:bzThermState}. Note that the terms where $\lambda$ appears have been truncated to $O(\lambda^{2})$. The term $\langle \langle \hat J_{x}^{2}+J_{y}^{2} \rangle \rangle$ is unaffected by noise as it commutes through the Hamiltonian. Therefore, for the spin variances, only the term $\Delta\langle\langle\hat{J}_{\rm{x}}\rangle\rangle^{2}$ comes into play; starting in $\hat J_{x}$ eigenstate means that, to $O(\lambda^{2})$, there is no noise contribution from $\langle\langle\hat{J}_{\rm{y}}\rangle\rangle$ either.

Using the previous bound in Eq. \ref{eq:EWboundThermState}, we plot the EW violation in Figure \ref{fig:AllThermPlots} (dotted lines). For violation to take place, the condition that 
\begin{equation}
   \ \frac{Q}{\bar n}> \frac{\pi [7+12 \bar n (1+\bar n)]}{6(1+2 \bar n)} \sim \pi \bar n
    \label{eq:QnbarRelationStartTherm}
\end{equation}
is required when $t=\pi/\omega$. The bound is approximately $\pi \bar n$ for $\bar n \gg 1$.

We can further consider the scenario where the oscillator is cooled to its ground state at the start, but remains in contact with a bath for the rest of time; this is given by $W_{\rm{en,0}, noise}=W_{\rm{en,0}}+\Delta W_{\rm{\bar n,noise}}$, with the coefficients of Eqs. \ref{eq:ay}-\ref{eq:bz} being used in Eq. \ref{eq:EWbathNoise} (see Supplemental Material for ``re-optimized" coefficients). The dot-dashed lines in Fig. \ref{fig:AllThermPlots} shows the violation for different values of $\bar n/Q$. The condition for violation to take place is now
\begin{equation}
   \ \frac{Q}{\bar n}> \frac{\csc(\frac{\omega t}{2})^{2}}{12}[10 \omega t-4 \omega t \cos(\omega t)- 4 \sin(\omega t)-\sin(2 \omega t)]
    \label{eq:QnbarRelationStartGround}
\end{equation}
which for $t = \pi/\omega$ is $\frac{Q}{\bar n}>(7 \pi/6)$. This is $O(\bar n)$ better than for an initial thermal state.



We now investigate the value of the magnetic coupling in Figure \ref{fig:OscillatorGeoPRL}. A large diamagnetic coupling can be attained by using superconducting masses and placing the atoms in different magnetically sensitive states. For instance, we can place one atom in the $m_{F}=1$ state and the other in $m_{F}=-1$. In such a scenario, the coupling would primarily be due to the first order Zeeman shift, which is linear in $B$  \cite{SteckCesium}. For a cylinder with $2R=L=5$ mm, $\chi = -1$, density$=11.34$ g/cm$^{3}$ \cite{Lead}, $\Delta z = 5~\mu$m, $s = 8$ mm, $R_{0}= 4.5$ cm, $B_{\rm{ext}}=$ 50 mT, $\omega = 2 \pi/20$, and $N=10^{6}$, we get $\lambda = 6.3 \times 10^{-4}$. A coupling approximately $\times 10^{2}$ weaker is available for atomic clock ($m_{F}=0$) states. For details on this calculation, see the Supplemental Material; this includes the cylindrical field expressions \cite{CylinderMagnetization}, and calculations for regular metals like tungsten \cite{Tungsten,TungstenSusc}.

\textbf{\textit{Conclusion -}} Within the context of a combined atom interferometer and oscillator system, we can use an entanglement witness to show that entanglement exists even in the case where the oscillator starts in a thermal state and remains in contact with a bath. In fact, we find that for the appropriate oscillator quality factor $Q$ and thermal occupancy $\Bar{n}$, violation always exists for any value of magnetic coupling $\lambda$. Additionally, an $O(\bar n)$ improvement in the EW violation if the oscillator is cooled to its ground state before running the experiment. However, to observe this violation, the number of measurements required would be \textit{at least} on the order of 10$^5$. Hence, experimentally using the proposed EW to measure entanglement may not be ideal; rather our results tell us that with the right physical parameters we can indeed experimentally generate entanglement between an atom interferometer and a mechanical oscillator, even with thermal noise. For practical detection of entanglement, other methods will have to be employed.

\textbf{\textit{Acknowledgments -}} We would like to acknowledge Yuxin Wang and Jon Kunjummen for  conceptual and mathematical discussions. We would also like to thank our collaborators for helping with experimental details and overall discussions, in particular, Jon Pratt, Garrett Louie, Cristian Panda, Prabudhya Bhattacharyya, Matt Tao, Lorenz Keck, James Egelhoff, John Manley, Stephan Schlamminger, Holger M\"uller, and Daniel Carney. We thank Daniel Barker and John Manley for comments on our paper. The work of G. P. and D. B. was made possible
by the Heising-Simons Foundation grant 2023-4467 ``Testing the Quantum Coherence of Gravity'' and  through the support of Grant 63121 from the John Templeton Foundation. The opinions expressed in this publication are those of the authors and do not necessarily reflect the views of the John Templeton Foundation. J. T. is solely funded by the National Institute of Standards and Technology.

\section{Supplemental Material: Entanglement witness for combined atom interferometer-mechanical oscillator setup}

\subsection{Entanglement witness}
\label{EntanglementWitness}

This section provides more details and discussion on the derivation of the entanglement witness and its violation in the noiseless and thermal noise scenarios. Additionally, we add a section on the effects of atomic dephasing.

\subsubsection{General entanglement witness form and bound}

We investigate the entanglement witness,
\begin{equation}
    W = \textrm{Var}(\hat{J}_{x})+\textrm{Var}(\hat{J}_{y}+a_{y}\hat{q}+b_{y}\hat{p})+\textrm{Var}(\hat{J}_{z}+a_{z}\hat{q}+b_{z}\hat{p}).
    \label{eq:EWmainSupp}
\end{equation}

As we stated in the main paper, the bound for $W$ can be derived using the theorem in the paper by Hofmann et al.~\cite{EWboundMain}: for operators $\hat{A}_{\mu}$ of system $A$ and $\hat{B}_{\mu}$ of system $B$, if $\sum_{\mu}\textrm{Var}(\hat{A}_{\mu}) \geq U_{A}$ and $\sum_{\mu}\textrm{Var}(\hat{B}_{\mu}) \geq U_{B}$, then $\sum_{\mu}\rm{Var}(\hat{A}_{\mu}+\hat{B}_{\mu}) \geq U_{A}+U_{B}$ for any separable state. There is, however, a prerequisite that for each operator $\hat{A}_{\mu}$ there must be a $\hat{B}_{\mu}$. In the case of Eq. \ref{eq:EWmainSupp}, $\hat{A}_{\mu} = \hat{J}_{\mu}$ and $\hat{B}_{\mu} = a_{\mu}\hat{q}+b_{\mu}\hat{p}$. It can be seen in Eq. \ref{eq:EWmainSupp} that there is a term with no oscillator operators, and composed of only the spin operator $\hat{J}_{x}$; we do not measure correlations between $\hat J_{x}$ and the oscillator operators because the mean of these correlations are zero, at least for the noiseless entangled state in Equation \ref{eq:EntState}. In order to satisfy the requirement of every $\hat{A}_{\mu}$ operator having its counterpart $\hat{B}_{\mu}$, we started with the term $\textrm{Var}(\hat{J}_{x}+a_{x}\hat{q}+b_{x}\hat{p})$ and set the coefficients $a_{x}$ and $b_{x}$ to zero \textit{after} the derivation of the bound. We lay out this derivation here.

Starting with $\tilde{W}=\sum_{\mu}\textrm{Var}(\hat{J}_{\mu}+a_{\mu}\hat{q}+b_{\mu}\hat{p})$, where the summation $\sum_{\mu}$ is over the $x,y,$ and $z$ components, we follow Ref.~\cite{EWboundMain} to arrive at the general bound
\begin{multline}
    \tilde{W} \geq \left[\sum_{\mu}\textrm{Var}(\hat{J}_{\mu})\right]_{\rm{min}}
    +\left[\sum_{\mu}\textrm{Var}(a_{\mu}\hat{q}+b_{\mu}\hat{p})\right]_{\rm{min}}.
\end{multline}
To obtain the witness in Eq. \ref{eq:EWmainSupp}, we set $a_{x}=0$ and $b_{x}=0$ on the left and right hand sides of the inequality.
The oscillator-operator bound can be derived as follows: we have
\begin{equation}
    \textrm{Var}(A)+\textrm{Var}(B) = \textrm{Var}(A)+\frac{\textrm{Var}(A)\textrm{Var}(B)}{\textrm{Var}(A)}
\end{equation}
for which we can apply the Heisenberg uncertain principle
\begin{equation}
    \textrm{Var}(A)\textrm{Var}(B) \geq \frac{1}{4}|\langle[A,B]\rangle|^{2}
\end{equation}
and minimize the R.H.S to give
\begin{equation}
    \textrm{Var}(A)+\textrm{Var}(B) \geq |\langle[A,B]\rangle|. 
\end{equation}
Setting $\hat{A}=a_{y} \hat q+b_{y} \hat p$ and $\hat{B}=a_{z} \hat q+b_{z} \hat p$, we then find that the sum of the oscillator-operator variances is bounded by
\begin{equation}
\sum_{\mu \neq x}\textrm{Var}(a_{\mu}\hat{q}+b_{\mu}\hat{p}) \geq |a_{y}b_{z}-a_{z}b_{y}|.
\label{OrthoXPboundSupp}
\end{equation}
In analogy to a derivation in Ref.~\cite{EWboundMain}, the general bound for the spin operators can be obtained as follows:
\begin{equation}
\sum_{\mu}\textrm{Var}(\hat{J}_{\mu}) = \sum_{\mu} (\langle \hat{J}_{\mu}^{2} \rangle  - \langle \hat{J}_{\mu} \rangle^{2})
\end{equation}
With $\sum_{\mu} \langle \hat{J}_{\mu}^{2} \rangle = j(j+1)$ and $\sum_{\mu} \langle \hat{J}_{\mu}\rangle^{2} \leq j^{2} $ we have
\begin{equation}
\sum_{\mu}\textrm{Var}(\hat{J}_{\mu}) \geq j.
\end{equation}
The question now arises as to the exact value of $j$. For a fully symmetric state, we have
\begin{equation}
    \sum_{\mu}\textrm{Var}(\hat{J}_{\mu}) \geq \frac{N}{2}.
    \label{eq:FullSymmBoundSupp}
\end{equation}
Here $j$ takes the maximum possible value. Eq. \ref{eq:FullSymmBoundSupp} can also be derived from the fact that $\sum_{\mu}[\textrm{Var}(\hat{\sigma}_{\mu})] \geq 2$ (shown in Ref.~\cite{EWboundMain}),
\begin{equation}
\left[\sum_{\mu}\textrm{Var}(\hat{J}_{\mu})\right]_{\textrm{min}} = N\left[\sum_{\mu}\textrm{Var}(\hat{S}_{\mu})\right]_{\textrm{min}}=\frac{N}{2}
\end{equation}

Using $[j]_{\textrm{min,exp}}$, the minimum value that $j$ will take at the time of measurement, we write the most general bound to be
\begin{equation}
    W_{b}= [j]_{\textrm{min,exp}}+|a_{y}b_{z}-a_{z}b_{y}|.
\end{equation} 
as in the main paper.

As mentioned, we consider the quantity 
\begin{equation}
    W_{\rm{ratio}}=\left[\frac{W_{b}-W_{\rm{en}}}{W_{\rm{b}}}\right]_{\rm{max}};
    \label{eq:EWratioSupp}
\end{equation} 
when $|W_{b}-W_{\rm{en}}| \ll |W_{\rm{b}}|$, finding $|W_{b}-W_{\rm{en}}|_{\rm{max}}$ suffices. Practically, however, finding $|W_{b}-W_{\rm{en}}|_{\rm{max}}$ is complicated due to the absolute value term given by $|a_{y}b_{z}-a_{z}b_{y}|$. Therefore, in this paper we minimize the value $W_{\rm{en}}$; that is, we solve for $a_{\mu}$ and $b_{\mu}$ such that $\partial_{a_{\mu}}(W_{\rm{en}}) = 0$ and $\partial_{b_{\mu}}(W_{\rm{en}}) = 0$. $W_{\rm{en}}$, being quadratic in $a_{\mu}$ and $b_{\mu}$, is solvable. Note that this gives a higher violation than if were to maximize $W_{b}-W_{\rm{en}}$ by expressing the absolute value as $s (a_{y}b_{z}-a_{z}b_{y})$, where $s$ sets the sign (+1, or -1), for reasons we cannot explain currently; the underlying reason must be that optimizing $s (a_{y}b_{z}-a_{z}b_{y})$ (and choosing the pairing of $s$ and the coefficients which makes this value positive) is not equivalent to optimizing $\sqrt{(a_{y}b_{z}-a_{z}b_{y})^{2}}$.

$a_{\mu}$ and $b_{\mu}$ contribute to obtaining $W_{\rm{en,min}}$ by reducing the value of the variance terms in Equation \ref{eq:EWmainSupp}. In the most general case, an expansion of Eq. \ref{eq:EWmainSupp} gives
{\small
\begin{multline}
    \ W = \sum_{\mu=x,y,z} \text{Var}(\hat J_{\mu}) 
+ \sum_{\mu=y,z}( a_{\mu}^{2} \langle \hat q^{2} \rangle+ b_{\mu} \langle \hat p^{2} \rangle + a_{\mu} \langle \hat J_{\mu} \hat q + \hat q \hat J_{\mu} \rangle \\
~~~~+ b_{\mu} \langle \hat J_{\mu} \hat p+ \hat p \hat J_{\mu}  \rangle 
+a_{\mu}b_{\mu}\{\hat q, \hat p\} - 2\langle a_{\mu} \hat q +b_{\mu} \hat p \rangle \langle \hat J_{\mu} \rangle - \langle a_{\mu} \hat q +b_{\mu} \hat p \rangle^{2})
    \label{eq:EWmainExpandSupp}
\end{multline}}
To obtain $W_{\rm{en}}$, the expectation value indicated by $\langle  \rangle$ is taken by evaluating the operators between the appropriate entangled state denoted by $\ket {\psi_{\rm{en}}}$. Observing the terms of $W$, we can see that $a_{\mu}$ and $b_{\mu}$ characterize the contribution of covariances to the value of the witness; for instance, there are terms such as $a_{\mu} \langle \hat J_{\mu} \hat q + \hat q \hat J_{\mu} \rangle$ and $b_{\mu} \langle \hat J_{\mu} \hat p+ \hat p \hat J_{\mu}  \rangle$. Additionally, as described in the upcoming sections, the spin and oscillator operators get correlated with each other, which can be seen when investigating their evolution in the Heisenberg picture. Therefore, in optimizing for the values of $a_{\mu}$ and $b_{\mu}$, we are primarily controlling the contribution of these covariances  and correlations, such that we obtain $W_{\rm{en,min}}$.

\subsubsection{Noiseless entanglement witness}
\label{Noiseless}

The noiseless entangled state of our system is given by \cite{PRXQuantum.2.030330}
\begin{equation}
    \ket{\psi_{\rm{en}}}= D^{\dagger}(\lambda \hat{J}_{z})e^{-i\omega \hat{c}^\dagger \hat{c} t}D(\lambda \hat{J}_{z})\ket{+++~...}\ket{0}.
    \label{eq:EntState}
\end{equation}
Compared with Ref.~\cite{PRXQuantum.2.030330}, we have dropped the $\exp[-i(g^{2}t/\omega)\hat{J}_{z}^{2}]$ squeezing term, because it's likely to be of $O(1)$ given the size of the coupling $g$.

As explained in the main paper, the Heisenberg operator $\hat{c}(t)$ is given by,
\begin{equation}
    \ \hat c(t) = e^{-i\omega t} \hat c(0)-i\int_{0}^{t} dt'~ g\hat J_{z}e^{-i\omega (t-t')}.
    \label{eq:loweringOpTime}
\end{equation}
The spin operators $\hat{J}_{x}$ and $\hat{J}_{y}$ rotate about the z-axis by angle $g \int_{0}^{t}dt'~ [\hat c(t')+\hat c(t')^\dagger]$, due to the term $g \hat J_{z}(\hat c + \hat c^\dagger)$ in the Hamiltonian. We find the operator evolution of $\hat{J}_{x}, \hat{J}_{y},\hat{q}$ and $\hat{p}$ to be:
\begin{align}
    \hat{J}_{x}(t) & \approx \hat{J}_{x}\cos(\theta_{\hat{p}}+\theta_{\hat{q}})-\hat{J}_{y}\sin(\theta_{\hat{p}}+\theta_{\hat{q}})
    \label{eq:JxHeis}\\
    \hat{J}_{y}(t) & \approx \hat{J}_{y}\cos(\theta_{\hat{p}}+\theta_{\hat{q}})+\hat{J}_{x}\sin(\theta_{\hat{p}}+\theta_{\hat{q}})
    \label{eq:JyHeis}\\
    \hat{q}(t) & = \hat{q} \cos(\omega t)+\frac{\hat{p}}{m \omega} \sin(\omega t)-\sqrt{\frac{2}{m \omega}} \lambda \hat{J}_{z}[1-\cos(\omega t)]
    \label{eq:qHeis}\\
     \hat{p}(t) & =\hat{p} \cos(\omega t)-m \omega \hat{q} \sin(\omega t)-\sqrt{2 m \omega} \lambda \hat{J}_{z}\sin(\omega t)
    \label{eq:pHeis}
\end{align}
where
\begin{align}  \theta_{\hat{p}}&=\sqrt{\frac{2}{m \omega}}\lambda \hat{p}[1-\cos(\omega t)] \\    
    \theta_{\hat{q}}&= \sqrt{2 m \omega} \lambda \hat{q} \sin(\omega t)
\end{align}
Note that we have dropped the squeezing term, which goes as $\lambda^{2} \hat{J}_{z}$ inside the cosine and sine functions of Equations \ref{eq:JxHeis} and \ref{eq:JyHeis}. As observed in these equations, the operators become correlated with each other (for instance, $\hat q(t)$ and $\hat J_{z}$). Optimizing the values of $a_{\mu}$ and $b_{\mu}$ translates to controlling the contributions of these correlations such that the variances of Eq. \ref{eq:EWmainSupp} are minimized for the entangled state. As described in the main paper, if we consider the operator combination $\hat{J}_{y}(t)+a_{y}\hat{q}(t)+b_{y}\hat{p}(t)$, $a_{y}$ and $b_{y}$ can be chosen such that the contribution to $\hat{J}_{y}(t)$ from $\hat{J}_{x}$ can be canceled. 

The non-simplified versions of these coefficients are 
\begin{align}
    a_{\rm{y,0}} &= -\frac{\sqrt{2m\omega} \lambda N \sin(\omega t)}{\{2+3\lambda^{2}N+\lambda^{2}N[\cos(2 \omega t)-4\cos(\omega t)]\}}
    \label{eq:ayNoiseless}\\
    b_{\rm{y,0}}&=\sqrt{\frac{2}{m\omega}}\frac{\lambda N [1-\cos(\omega t)]}{[2+\lambda^{2}N-\lambda^{2}N\cos(2\omega t)]}\\
    a_{\rm{z,0}} &= \frac{\sqrt{2m\omega} \lambda N [1-\cos(\omega t)]}{\{2+3\lambda^{2}N+\lambda^{2}N[\cos(2 \omega t)-4\cos(\omega t)]\}}
    \label{eq:azNoiseless}\\
   b_{\rm{z,0}}&=\sqrt{\frac{2}{m\omega}}\frac{\lambda N \sin(\omega t)}{[2+\lambda^{2}N-\lambda^{2}N\cos(2\omega t)]}
\end{align}
which, when written to $O(\lambda)$, give Eqs. \ref{eq:ay}-\ref{eq:bz} in the main paper.

The lowest order of the expectation values $\langle \hat{J}_{x} \rangle$, $\langle \hat{J}_{x}^{2} \rangle$, $\langle \hat{J}_{y} \rangle$, $\langle \hat{J}_{y}^{2} \rangle$, $\langle \hat{q}^{2} \rangle$, and $\langle \hat{p}^{2} \rangle$ is $\lambda^{2}$. $\langle \hat{q}\hat{J}_{x}+\hat{J}_{x}\hat{q}\rangle = \langle \hat{p}\hat{J}_{x}+\hat{J}_{x}\hat{p}\rangle = 0$. Very interestingly, to lowest order in $\lambda$, $\langle \hat{q}\hat{J}_{y}+\hat{J}_{y}\hat{q}\rangle = -a_{\rm{y,0,\lambda}}$, $\langle \hat{p}\hat{J}_{y}+\hat{J}_{y}\hat{p}\rangle = -a_{\rm{z,0,\lambda}}$, $\langle \hat{q}\hat{J}_{z}+\hat{J}_{z}\hat{q}\rangle = -b_{\rm{y,0,\lambda}}$, and $\langle \hat{p}\hat{J}_{z}+\hat{J}_{z}\hat{p}\rangle = -b_{\rm{z,0,\lambda}}$, where $a_{\rm{y,0,\lambda}}$ etc. are the expressions Eqs. \ref{eq:ay}-\ref{eq:bz} of the main paper. It must be noted that these coefficients reduced to being $O(\lambda)$ to lowest order in the end, despite the fact that the calculations were conducted to $O(\lambda^{2})$.

We can also understand the coefficients by investigating how the sum of the spin variances change as the system evolves. To $O(\lambda^{2})$, we find that
\begin{equation}
  \sum_{\mu}\textrm{Var}(\hat{J}_{\mu}) = \frac{N}{2}+\frac{\lambda^{2}N^{2}}{2}[1-\cos(\omega t)].
  \label{eq:SpinVarSum}
\end{equation}  
for $\ket{\psi_{\rm{en}}}$. This shows that there is overall increase in $\left[\sum_{\mu}\textrm{Var}(\hat{J}_{\mu})\right]$ as the system evolves. However, if we consider the contribution from the rest of the operators in $W$ (see Eq. \ref{eq:EWmainExpandSupp}), we find that the coefficients work with the $\hat q$ and $\hat p$ operators to bring the overall value of $W_{\rm{en}}$ to be less than $N/2$; this contribution is given by
\begin{equation}
  \  W_{\rm{en,0}} -\sum_{\mu}\textrm{Var}(\hat{J}_{\mu}) = -\lambda^{2}N^{2}[1-\cos(\omega t)],
   \label{eq:EWminVarJ}
\end{equation} 
such that the total value of $W_{\rm{en}}$ is
\begin{equation}
    W_{\rm{en,0}}=\frac{N}{2}-\frac{\lambda^{2}N^{2}}{2}[1-\cos(\omega t)].
    \label{eq:EWvalueEntStateSupp}
\end{equation}
In essence, the coefficients work in various ways with the terms in Eq. \ref{eq:EWmainExpandSupp} to minimize $W_{\rm{en}}$.

Note that $W_{\rm{en,0}}$ and $W_{\rm{b,0}}$ are obtained by minimizing the value of the EW for state $\ket{\psi}_{\rm{en}}$. While this means that Eq. \ref{eq:EWvalueEntStateSupp} is the lowest value for the EW that can be obtained with $\ket{\psi}_{\rm{en}}$, it may not necessarily provide us with the greatest difference between the separable-state bound and the entangled-state value, which would be obtained by maximizing the quantity $W_{\rm{b}}- W_{\rm{en}}$ with respect to $a_{\mu}$ and $b_{\mu}$. The reason that we have not presented results from the optimization of $W_{\rm{b}}- W_{\rm{en}}$ is due to certain mathematical complexities. 

Note also that, though we present the coefficients to $O(\lambda)$ in the main paper, when calculating $W_{b,0}$ and $W_{\rm{en},0}$, we used the non-simplified expressions given in this section.

If experimentally measuring the entanglement witness, measurements of \textit{all} the operators of a single variance term must be made in a single experimental run; that is, in a single run we must measure all the operators in the set $\{\hat{J}_{y},\hat{J}_{y}^{2},\hat{q}^{2},\hat{q}\hat{J}_{y}+\hat{J}_{y}\hat{q},\hat{p}\hat{J}_{y}+\hat{J}_{y}\hat{p}\}$, with analogous sets for the $\mu=x$ and $\mu=z$ variance terms.
After the completion of all runs, appropriate averaging over these runs is conducted to obtain $\langle \hat{J}_{x} \rangle$, $\langle \hat{J}_{x}^{2} \rangle$, etc.. Due to the existence of noise, these averages are not purely the expectation values of operators taken over a quantum state. We will also be averaging over noise, so it is in fact more accurate to state that the final experimental values we obtain are those for $\langle\langle~\langle \hat{J}_{x}  \rangle~\rangle\rangle$, $\langle\langle~\langle \hat{J}_{x}^{2}  \rangle~\rangle\rangle$ etc., where double-angle brackets denote averaging over classical noise. We look at the cases of atomic dephasing and thermal noise in the upcoming sections.

\subsubsection{Entanglement witness with atomic dephasing}

\begin{figure*}[htb]
         \includegraphics[width = \textwidth]{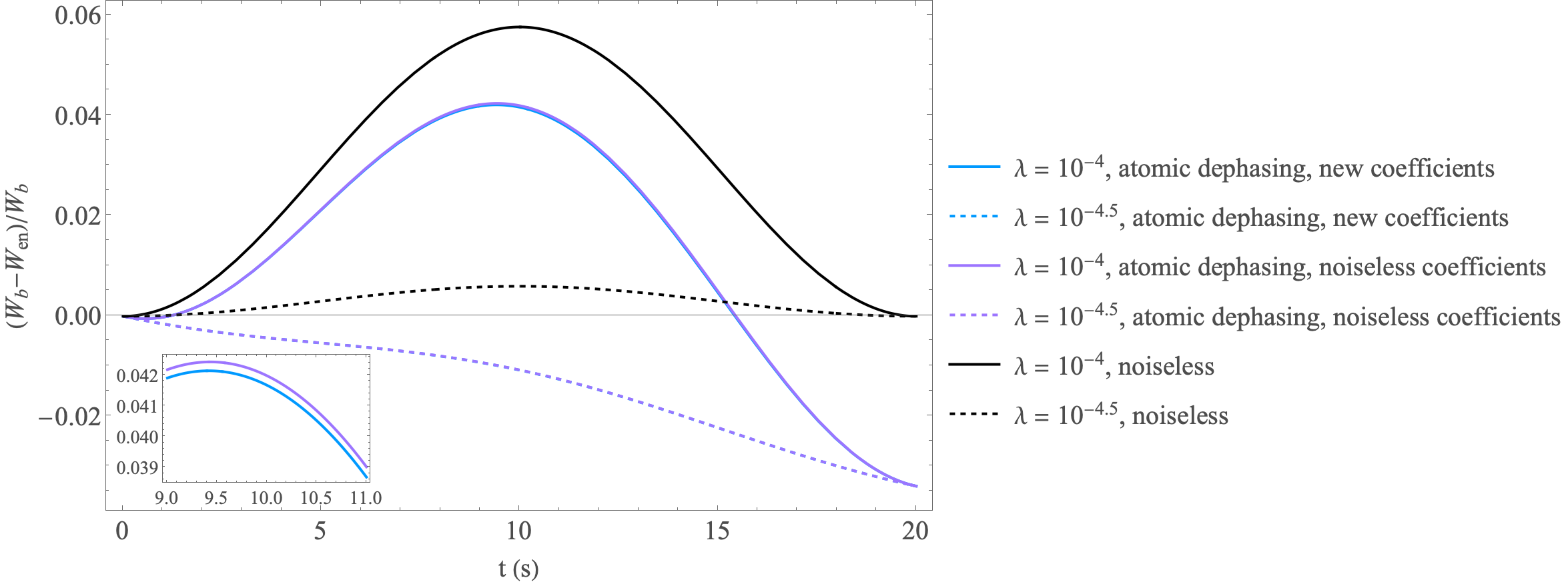}
       \caption{Here, $\omega = 2 \pi/20$ rad s$^{-1}$, $\sigma^{2} = t/600$, $N=10^{6}$, and we have set $[j]_{\rm{min,exp,start}} = N/2$. Note that the coupling satisfies $\lambda< 7 \times 10^{-4}$. As expected, dephasing reduces the violation.} 
       \label{fig:NoiselessPlusAtomDeph}
\end{figure*}

In reality, our experiment is subject to noise which can affect the entangled-state value of the EW. This can result in decreasing the value of $W_{\rm{b}}- W_{\rm{en}}$, potentially making it more difficult to observe violation of the entanglement witness.

Atomic dephasing can be introduced to the entangled state by introducing a spin-rotation $a_{j}$ about the z-axis for each atom $j$:
\begin{multline}
    \ket{\psi}_{\rm{en,\alpha_{j}}}= D^{\dagger}(\lambda J_{z})e^{-i\omega c^\dagger ct}D(\lambda J_{z}) \\
    \times \exp\left[i\sum_{j}\hat{S}_{z,j}\alpha_{j}\right]\ket{+++~...}\ket{0}
    \label{eq:EntStateDeph}
\end{multline}
We evaluate each of the expectation value terms in Eq. \ref{eq:EWmainExpandSupp} for $\ket{\psi}_{\rm{en,\alpha_{j}}}$; as previously, we evolve the operators in the Heisenberg picture first, with appropriate modifications made to Eqs. \ref{eq:JxHeis}-\ref{eq:pHeis} as dictated by the spin-rotation $\alpha_{j}$, after which we evaluate with respect to $\ket{++...}\ket{0}$. Taking each $\alpha_{j}$ to be independent of each other and having a Gaussian distribution of
\begin{equation}
   p(\alpha_{j})=\frac{1}{\sqrt{2\pi\sigma^{2}}}\exp\left[\frac{-\alpha_{j}^{2}}{2\sigma^{2}}\right],
\end{equation}
we obtain the classical mean of each of these terms (that is, expressions for $\langle\langle~ \langle \hat{J}_{x} \rangle~ \rangle\rangle$, $\langle\langle~ \langle \hat{J}_{x}^{2} \rangle~ \rangle\rangle$ etc.) with respect to the angles $\alpha_{j}$. The notation $\langle\langle~\langle... \rangle~\rangle\rangle$ refers to the process of first evaluating the expectation value of the operators, $\langle~\rangle$, and afterwards the classical mean, $\langle \langle~\rangle \rangle$.

Compared to the noiseless case, we have the following change $\Delta$ to the variances of the spin operators for $\ket{\psi_{\rm{en}}}$, to $O(\sigma^{2})$:
{\small
\begin{align}
   \Delta \textrm{Var}(\hat J_{x}) &= \left\{\frac{N}{4}-\frac{\lambda^{2}N}{2}[1-\cos(\omega t)]\right\}\sigma^{2}\\
   \Delta \textrm{Var}(\hat J_{y}) &= \left\{\frac{\lambda^{2}N}{2}[1-\cos(\omega t)]-\frac{\lambda^{2}N^{2}}{2}[1-\cos(\omega t)]\right\}\sigma^{2}
\end{align}}
Since $\hat{J}_{z}$ commutes with the dephasing operators $\hat{S}_{z,j}$, its variance remains intact. 

With these, the value of the EW with the inclusion of atomic dephasing is 
\begin{equation}
   \langle\langle W_{\rm{en,\sigma,non-opt}} \rangle\rangle =
   \frac{N}{2}\left(1+\frac{\sigma^{2}}{2}\right)-\frac{\lambda^{2}N^{2}}{2}[1-\cos(\omega t)].
   \label{eq:EWentStateDeph}
\end{equation}
Note that the expressions are to $O(\sigma^{2})$; $\sigma = \sqrt{t/\tau_c}$, where $\tau_c$ is the coherence time of interferometer which we require to be large for interferometry; therefore, we expect $\sigma \ll  1$. Here, we have used the same values of $a_{\mu}$ and $b_{\mu}$ as for the noiseless case. The $(N/4)\sigma^{2}$ difference between Equations \ref{eq:EWentStateDeph} and \ref{eq:EWvalueEntStateSupp} arises from dephasing-induced changes to the sum of the spin variances $\sum_{\mu}\textrm{Var}(\hat{J}_{\mu})$. If we break down the contributions from the different operators, we have
\begin{equation}
  \sum_{\mu}\textrm{Var}(\hat{J}_{\mu}) = \frac{N}{2}\left(1+\frac{\sigma^{2}}{2}\right)+\frac{\lambda^{2}N^{2}}{2}[1-\cos(\omega t)](1-\sigma^{2})
   \label{eq:VarSumDeph}
\end{equation}
while the rest of the operators give
{\small
\begin{equation}
  \ \langle\langle W_{\rm{en,\sigma,non-opt}} \rangle\rangle-\sum_{\mu}\textrm{Var}(\hat{J}_{\mu}) = -\lambda^{2}N^{2}[1-\cos(\omega t)]\left(1-\frac{\sigma^{2}}{2}\right)
   \label{eq:EWminVarJDeph}
\end{equation}}
It can be seen that $\langle\langle W_{\rm{en,\sigma,non-opt}} \rangle\rangle > W_{\rm{en,0}}$ (given in Eq. \ref{eq:EWvalueEntStateSupp}), showing that atomic dephasing can make it harder to violate the separable-state bound. Since we used the same coefficients as for the noiseless case, the EW bound is partially given by $W_{b,0}$ of the noiseless case,
{\small
\begin{equation}
    \langle\langle W_{\rm{b,\sigma,non-opt}} \rangle\rangle = [j]_{\textrm{min,exp,start}}\left(1-\frac{\sigma^{2}}{2}\right)+\lambda^{2}N^{2}|[1-\cos(\omega t)]|
    \label{eq:EWexpBoundDeph}
\end{equation}}
$[j]_{\rm{min,exp,start}}$ is the minimum value of the total spin variances for a separable state as can be experimentally achieved at the \textit{start} of the experiment; its occurrence in this equation, which reduces the violation by $\sigma^2/2$, indicates the value we expect it to reduce to just before the measurement as atomic dephasing reduces the total $j$ quantum number.

We can further re-optimize the coefficients $a_{\mu}$ and $b_{\mu}$ for the case of dephasing by applying the constrain that $a_{y,\sigma}=-(a_{z,\sigma}b_{z,\sigma})/b_{y,\sigma}$. To $O(\sigma^2 )$ and $O(\lambda)$ we have
\begin{align}
a_{y,\sigma}=  -\lambda N \sqrt{\frac{m \omega}{2}} \sin(
   \omega t) \frac{\{4 - \sigma^2[1 + \cos(\omega t)]\}}{4} \\
a_{z,\sigma}=\lambda N \sqrt{\frac{{m \omega}}{2}} [1-\cos(\omega t)]\frac{\{4 - \sigma^2[1 + \cos(\omega t)]\}}{4} \\
b_{y,\sigma}=\lambda N \sqrt{\frac{1}{2 m \omega}} [1-\cos(\omega t)] \frac{\{4 - \sigma^2[1 - \cos(\omega t)]\}}{4}\\
b_{z,\sigma}=\lambda N\sqrt{\frac{1}{2 m \omega}} \sin(
   \omega t)\frac{\{4 - \sigma^2[1 - \cos(\omega t)]\}}{4}
\end{align} 
Note that in the succeeding calculations, we used less simplified expressions; we do not present them here due to their complexity. As can be verified, these coefficients satisfy $\mathbf{a}.\mathbf{b}=0$. The bound is now
{\small
\begin{equation}
\begin{split}
    \langle\langle W_{\rm{b,\sigma}} \rangle\rangle &=[j]_{\textrm{min,exp,start}}\left(1-\frac{\sigma^{2}}{2}\right) \\ & +\lambda^{2}N^{2}[1-\cos(\omega t)]\left(1-\frac{\sigma^{2}}{2}\right).
    \label{eq:EWdephBoundOpt}
\end{split}    
\end{equation}}
The entangled-state value for the optimized coefficients remains the same as that of Eq. \ref{eq:EWentStateDeph} to $O(\lambda^{2})$ and $O(\sigma^{2})$; that is, the ``optimized" coefficients give the same entangled state value as when using noiseless coefficients (given in Eq. \ref{eq:EWentStateDeph}) to $O(\sigma^{2})$. Thus, we may as well use the noiseless coefficients for the case of atomic dephasing, because the ``optimized" coefficients only serve to decrease $\langle\langle W_{\rm{b,\sigma}} \rangle\rangle$ while keeping $\langle\langle W_{\rm{en,\sigma,non-opt}} \rangle\rangle$ unchanged.

Due to our initial expansions to $O(\lambda^{2})$, there is a maximum value that $\lambda$ can take such that Eq. \ref{eq:EWentStateDeph} is always positive; this condition is given by
\begin{equation}
   \lambda < \sqrt{\frac{\left(1+\frac{\sigma^{2}}{2}\right)}{2 N}}
    \label{eq:LambdaMaxDeph}
\end{equation}
For $\sigma = \sqrt{t/600}$, t = 20 s, and $N = 10^{6}$, it is required that $\lambda < 7 \times 10^{-4} $. We have plotted the violation in Fig. \ref{fig:NoiselessPlusAtomDeph}.  As expected, dephasing reduces the violation.

\subsubsection{Entanglement witness with a thermal state and white noise bath}

\begin{figure*}[htb]

    \includegraphics[width = \textwidth]
     {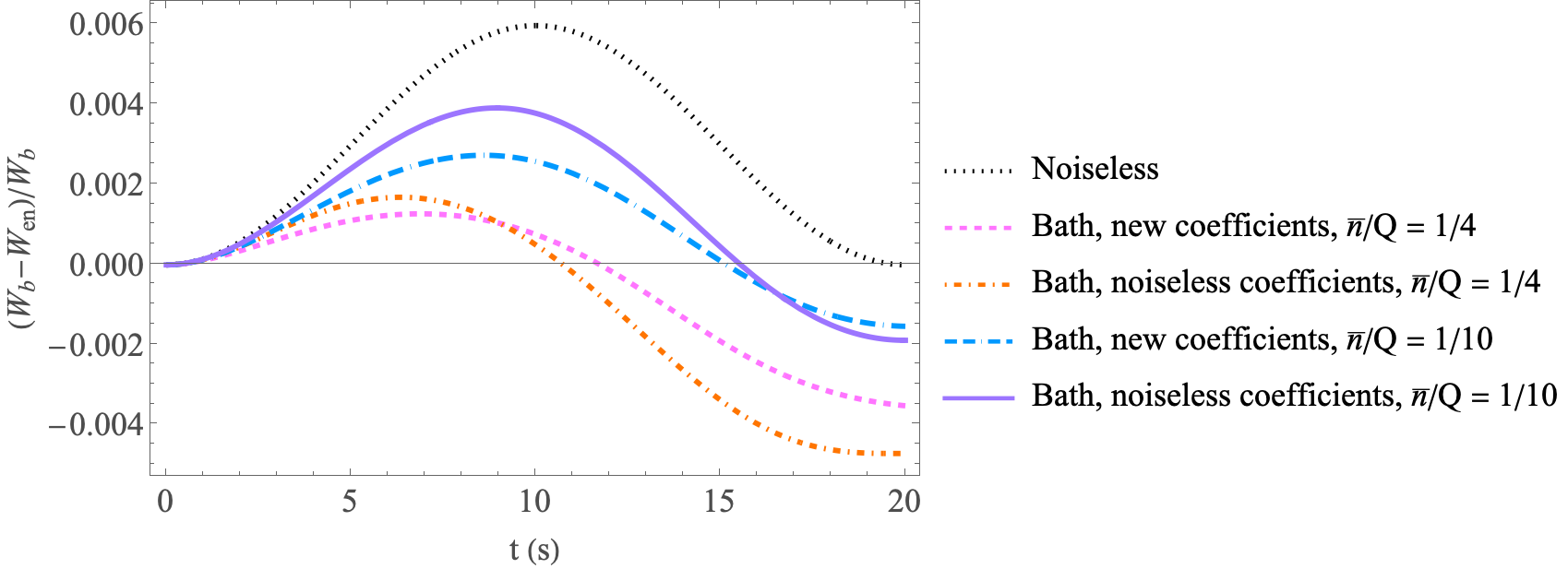}
       \caption{EW violation when the oscillator starts in a ground state and remains in contact with a thermal bath. $\lambda = 10^{-4.5}$, $N = 10^{6}$, and $\omega = 2 \pi/20$. We can see that re-optimized coefficients perform better for certain values of $t$ and  $\bar n/Q$, while the noiseless coefficients perform better for others.} 
       \label{fig:GroundStatePlusBath}
\end{figure*}

We now elaborate on the case of thermal noise. There are three scenarios that can be considered; the first when the oscillator starts in a thermal state, the second when the oscillator starts in a thermal state and remains in contact with a white noise bath, and the third when the oscillator starts in the \textit{ground} state and remains in contact with a bath.

For a thermal state \cite{ThermState}
\begin{equation}
    \ \Hat{\rho} = \sum_{n=0}^{\infty} \frac{e^{- n \beta \omega}}{1-e^{-  \beta \omega}} \ket{n}\bra{n}, 
    \label{eq:ThermState}
\end{equation}
introducing the initial-thermal-state-only effect into the previous noiseless calculation is rather straightforward. For average thermal occupancy $\Bar{n}$, we need only replace the expectation values of any any $\hat q^{2}$ and $\hat p^{2}$ terms that appear in the evaluation of the Heisenberg picture operators as follows
\begin{align}
 \langle\langle q^{2} \rangle\rangle: \frac{1}{2 m \omega} \rightarrow \frac{1}{2 m \omega} (2 \Bar{n} + 1)\\
\langle\langle p^{2} \rangle\rangle: \frac{m \omega}{2} \rightarrow \frac{m \omega}{2} (2 \Bar{n} + 1).
\end{align}

We can then find the coefficients as listed in the main paper, along with the bound $W_{\rm{b,\bar n}}$ and entangled-state value $W_{\rm{en,\bar n}}$. 

Next, we consider the scenario where the oscillator is in continuous contact with a white noise bath. As explained in the main paper, we can solve for $\hat{c}(t)$ to obtain
\begin{equation}
    \ \hat c(t) = e^{-i\omega t} \hat c(0)-i\int_{0}^{t} dt'~ (g\hat J_{z}+F_{in}(t'))e^{-i\omega (t-t')}.
    \label{eq:loweringOpTime}
\end{equation}
With $\hat q(t)=q_{\rm{zpf}}[\hat c(t) + \hat c^{\dagger}(t)]$ and $\hat p(t)=i p_{\rm{zpf}}[-\hat c(t) + \hat c^{\dagger}(t)]$, the noise contributions to $ \hat q(t)$ and $\hat p(t)$ can be expressed as
\begin{align}
    \ \mathcal{Q}(t) = -2 \sqrt{1/(2 m \omega)} \int_{0}^{t}dt'~ F_{\rm{in}}(t') \sin[\omega (t-t')]
    \label{eq:Q(t)}\\
    \ \mathcal{P}(t) = -2 \sqrt{m\omega/2} \int_{0}^{t}dt'~ F_{\rm{in}}(t') \cos[\omega (t-t')]
    \label{eq:P(t)}
\end{align}
Further, $\hat{J}_{x}$ and $\hat{J}_{y}$ will rotate about the z-axis by an angle $\xi(t) = g \int_{0}^{t}dt'~ [\hat c(t')+\hat c(t')^\dagger]$ as follows:
\begin{align}
    \hat{J}_{x}(t) = \hat{J}_{x} \cos[\xi(t)] - \hat{J}_{y} \sin[\xi(t)]\\
    \ \hat{J}_{y}(t) = \hat{J}_{y} \cos[\xi(t)] + \hat{J}_{x} \sin[\xi(t)]
    \label{eq:JyJxBathRot}
\end{align}
We can define the thermal noise contribution to this angle as
\begin{multline}
    \ \Xi(t) = -\sqrt{2 m \omega}~g \int_{0}^{t} dt'~ \mathcal{Q}(t')\\
 = 2 g \int_{0}^{t}\int_{0}^{t'} dt''dt'~F_{\rm{in}}(t'')\sin[\omega(t'-t'')]
    \label{eq:Xi}
\end{multline}

The covariances that contribute to the bath noise are as follows:
\begin{multline}
     \ \langle\langle \Xi(t)\Xi(t) \rangle\rangle = \frac{\lambda^{2}\Bar{n}}{Q}[6 \omega t - 8 \sin(\omega t)+\sin(2 \omega t)]\\
   \langle\langle \Xi(t)\mathcal{Q}(t) \rangle\rangle = -\sqrt{\frac{1}{2 m \omega}}\frac{8 \lambda \Bar{n}}{Q}\sin^{4}\left(\frac{\omega t}{2}\right)\\
     \langle\langle \Xi(t)\mathcal{P}(t) \rangle\rangle = \sqrt{\frac{m \omega}{2}}\frac{4 \lambda \Bar{n}}{Q}\left[\frac{\omega t}{2}-\sin(\omega t)+\frac{\sin(2\omega t)}{4}\right]\\
    \langle\langle \mathcal{Q}(t)\mathcal{Q}(t) \rangle\rangle = \frac{1}{2 m \omega}\frac{\Bar{n}}{Q}[2\omega t -\sin(2 \omega t)]\\
     \langle\langle \mathcal{P}(t)\mathcal{P}(t) \rangle\rangle = \frac{m \omega}{2}\frac{\Bar{n}}{Q}[2\omega t +\sin(2 \omega t)]\\
     \langle\langle \mathcal{Q}(t)\mathcal{P}(t) \rangle\rangle = \frac{\Bar{n}}{Q}\sin^{2}(\omega t).
     \label{eq:Covariances}
\end{multline}
As examples, $\langle\langle \Xi(t)\Xi(t) \rangle\rangle$ contributes to the noise in the EW term $\textrm{Var}(\hat J_{x})$ and $\langle\langle \mathcal{Q}(t)\mathcal{Q}(t) \rangle\rangle$ contributes to the noise in $\langle\langle \hat q^{2} \rangle\rangle$. Therefore, the expectation values that compose the EW must be modified as shown in the main paper.

Next, we move to the scenario where the oscillator starts in a ground state and remains in contact with a white noise bath. Some additional plots to those in the main paper are shown in Fig. \ref{fig:GroundStatePlusBath}; in particular, plots for ``re-optimized" coefficients. It can be seen that for certain values of $\bar n/Q$, the re-optimized coefficients show higher violation for certain times than the noiseless coefficients, while for other values of $\bar n/Q$, the noiseless coefficients are better.

\begin{figure*}[htb]
         \includegraphics[width = 15 cm]{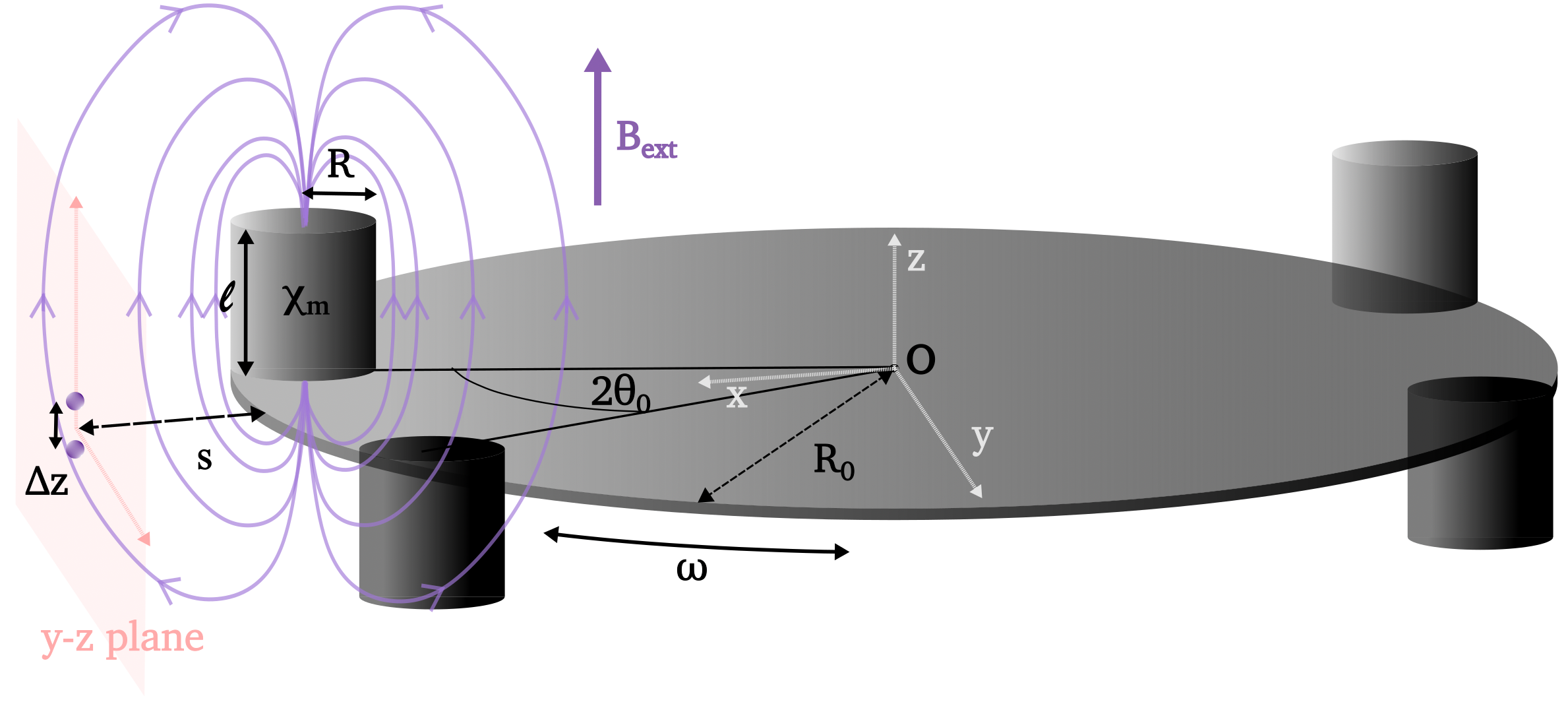}
       \caption{More advanced version of oscillator-atom setup. The parameter definitions are given in Table \ref{tab:PhysLimitsParameters}. An external magnetic field $\mathbf{B_{\rm{ext}}}$ induces a magnetic field in the cylinders which have magnetic susceptibility $\chi_{m}$. Note that, while only the field lines for one of the cylinders are shown for presentation purposes, both cylinders next to the atoms will be magnetized in reality. The direction of the induced magnetic field depends on the sign of $\chi_{m}$. The oscillator rotates in the x-y plane with frequency $\omega$. The cylinders are separated by angle $2\theta_{0}$. The reason for this configuration is to translate a force along the z-direction (as applied to the atoms) to a torque in the x-y plane. The two cylinders on the right are to balance the oscillator.} 
       \label{fig:OscillatorGeo}
\end{figure*}

 \begin{figure*}
    \centering
    \subfloat[]{
         \includegraphics[width=8.6 cm]{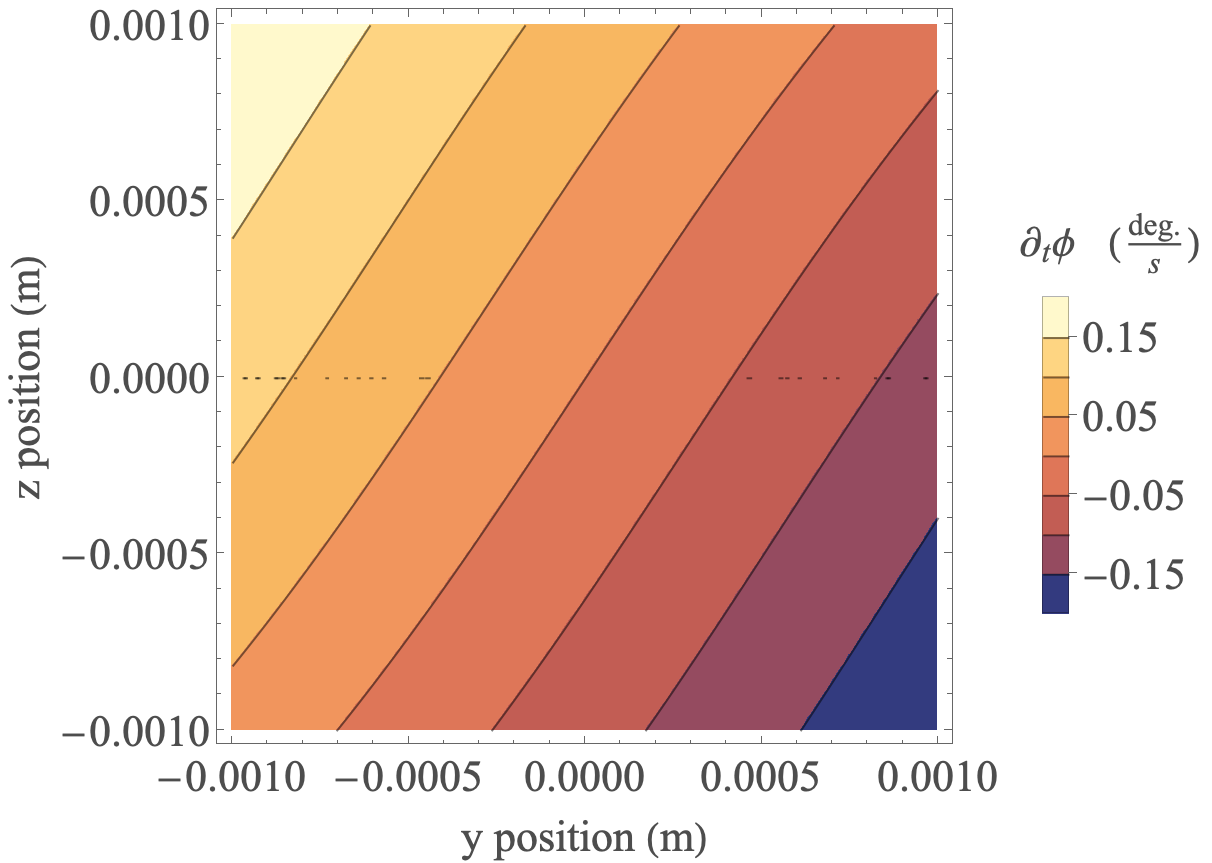}
         \label{fig:phidot_yz}
        }
     \hfill
     \subfloat[]{
         \includegraphics[width=8.6 cm]{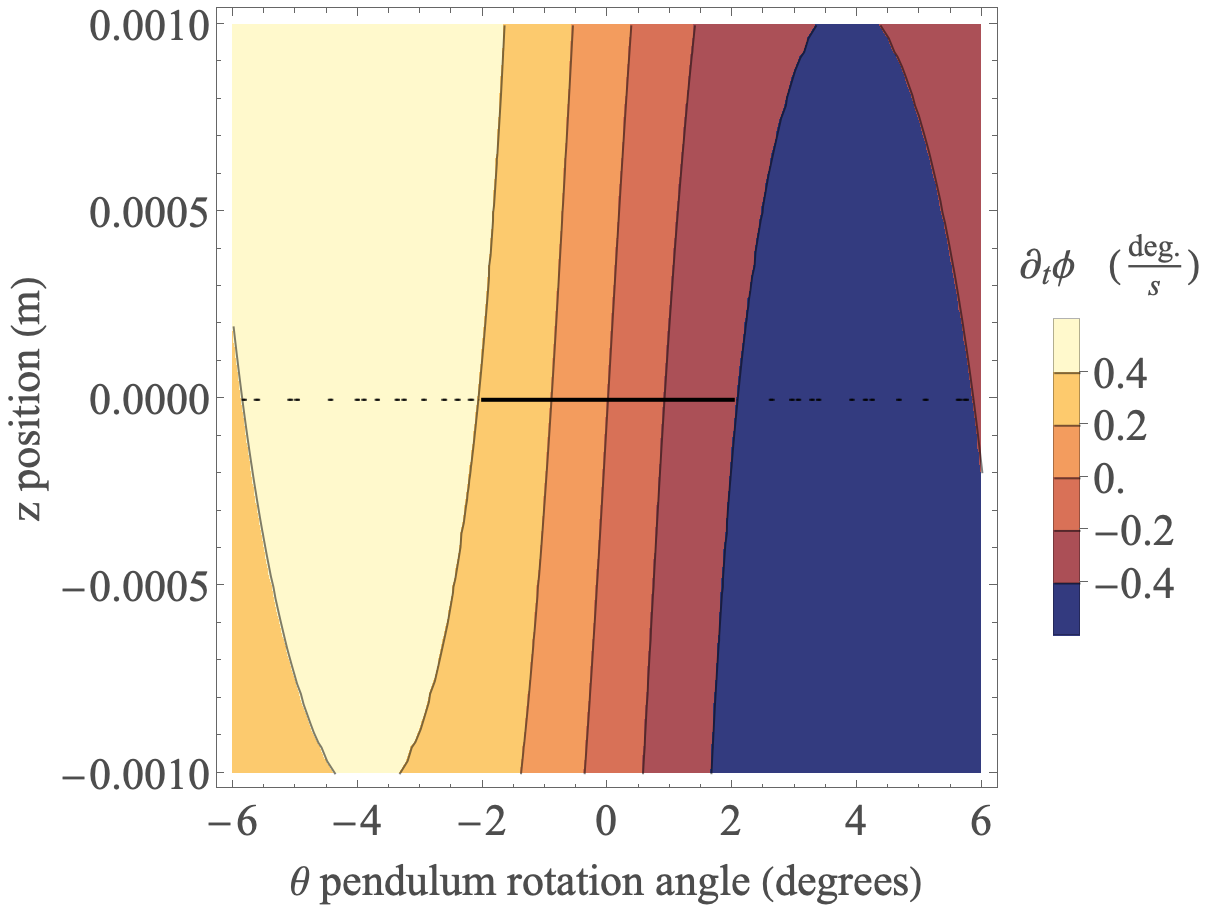}
         \label{fig:phidot_thetaz}
      }
        \caption{The physical parameters used are given in Table \ref{tab:PhysLimitsParameters}. In (a), the interferometer phase accumulation rate is plotted for different locations of the atoms in the y-z plane shown in Figure \ref{fig:OscillatorGeo}. In (b), the phase accumulation rate is plotted for different z locations of the atoms and rotation angles of the oscillator. The black line is the path the atom oscillator system follows to produce the interferometer phase plotted in Figure \ref{fig:simulatedphase}.}
        \label{fig:AccelTrapDynamics}
\end{figure*}

\subsection{Magnetic interactions}
\label{MagneticInteractions}

We now explain in detail the interaction between an atom interferometer and a mechanical oscillator, and how to the obtain the magnetic coupling between them. 

Atom interferometers measure the phase difference accumulated by the components of an atom's wave function that travel along different trajectories between two points in spacetime. In general, the phase accumulated by each component is the integral of the Lagrangian along the classical trajectory it traverses. In an optical lattice atom interferometer, the trajectories include a hold period where the atom is in a superposition of being trapped at different sites in an optical lattice. When this hold period is much longer than the time for the trajectories to separate and recombine, the dominant contribution to the relative phase is the integral of the force, $F_a(t)$, across the atoms at $r_a$ through the hold time, $T$, multiplied by the atom separation $\Delta r_a$. In units with $\hbar = 1$, the interferometer phase has the following form \cite{AtomInterExpHolger}:
\begin{equation}
    \Delta \phi = \Delta r_a \int_0^T dt F_a(t)
    \label{phase_equation}
\end{equation} 

\subsubsection{Interactions through the quadratic Zeeman effect}
If the atoms are exposed to a spatially varying magnetic field, then a force across the atoms is generated due to the gradient in the Zeeman splitting of the atom's energy levels. For Cesium-133 atoms in the $m_F = 0$ ground state, the leading order Zeeman effect is quadratic in the magnetic field \cite{SteckCesium}, yielding a force  
\begin{equation}
    F_a \propto \frac{\partial B^2(r_a)}{\partial r_a}. 
\end{equation}
Thus, atom interferometers can act as magnetometers, measuring the gradient in the magnitude of the magnetic field. If the source of the magnetic field is an oscillator composed of magnetic material, then the force across the atoms will encode the displacement of the oscillator, $r_o$. 
\begin{equation}
    F_a(r_o)  \propto \frac{\partial B^2(r_o,r_a)}{\partial r_a} 
\end{equation}
The atoms will feel a force that is modulated by the oscillator's motion. This results in an interferometer phase that is sensitive to the oscillators position. There will also be a back action force on the oscillator, $F_o$, from the dependence of the magnetic field on the oscillator's position. This force will be sensitive to the atom's position.
\begin{equation}
    F_o(r_a)  \propto \frac{\partial B^2(r_o,r_a)}{\partial r_o} 
\end{equation}
This physics is captured by the interaction potential for small displacements in the atom and oscillator coordinates.
 \begin{equation}
      U(r_o, r_a) = \phi_0 \frac{\partial^2B^2( 0, 0)}{\partial r_o \partial r_a} r_a  r_o
    \label{eq:interaction_potential}
 \end{equation}
The oscillator motion is now subject to a linear potential which modifies its motion and the potential interacts with the atoms, spatially separated by $\Delta r_a$, to accumulate interferometer phase at a rate $\Dot \phi$,
 \begin{equation}
      \Dot{\phi} = \phi_0 \dfrac{\partial^2B^2( 0, 0)}{\partial r_o \partial r_a} r_o \Delta r_a
 \end{equation}
Here, $\phi_0 $ is an atom's energy difference between the $m_{F}=0$ states of the ground state. To understand the magnitude of this interaction achievable with current technologies and how it can be increased, we model a particular experimental setup in the next section.

\subsubsection{Proposed setup}
\begin{figure}[htb]
         \includegraphics[width = 8.6 cm]{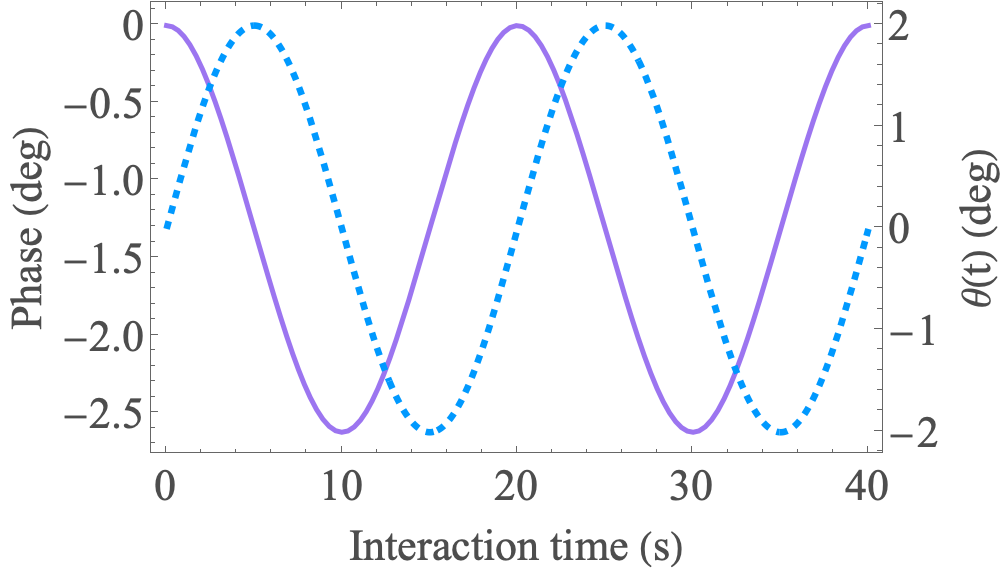}
       \caption{The physical parameters used are given in Table \ref{tab:PhysLimitsParameters}. The interferometer phase from magnetic interaction with the oscillator in classical motion is plotted with a solid line. The rotation angle of the oscillator is plotted in a dashed line. The oscillator motion is clearly encoded in the interferometer phase.} 
       \label{fig:simulatedphase}
\end{figure}

\begin{table}[htb]
\caption{\label{tab:PhysLimitsParameters}
The model of the atom-oscillator system has the following parameters, unless otherwise stated.
}
\begin{ruledtabular}
\begin{tabular}{ccc}
\textrm{Parameter}&
\textrm{Symbol}&
\textrm{Value}\\
\colrule
$m_F=0$ Cs ground state shift \cite{SteckCesium} & $\phi_0$ & 268.575 $(\frac{\text{rad}}{\text{ns}\text{ T}^2})$\\
External magnetic field strength & $B_{\rm{ext}}$ & $5$ mT \\
magnetic susceptibility \cite{TungstenSusc} & $\chi_m$ & $6.8 \times 10^{-5}$\\
pendulum radius & $R_0$ & 10 cm \\
pendulum frequency & $\omega$ & $2\pi/20$ rad/s \\
pendulum density \cite{Tungsten} & $\rho_{\rm{cyl}}$ & 19.3 g/cm$^{3}$ \\
cylinder height & $l$ & 10 mm \\
cylinder radius & $R$ & $5$ mm \\
half of cylinder separation angle & $\theta_0$ & $2^{\circ}$ \\
atom to pendulum distance & $s$ & 8 mm \\
atom wave packet separation & $\Delta z$ & 5 $\mu$m \\
\end{tabular}
\end{ruledtabular}
\end{table}

The main paper proposed a simplified version of the torsional pendulum. Since we plan to gradually proceed to a gravity experiment, we investigated a more advanced version of a pendulum as depicted in Figure \ref{fig:OscillatorGeo}. The pendulum that will oscillate at frequency $\omega$ about the vertical axis with two diamagnetic cylinders with magnetic susceptibility $\chi_m$ on its base. The atom interferometer sequence will trap atoms in a superposition of two lattice sites that are horizontally separated from the pendulum base by a distance, $s$, and vertically separated above and below the base by $\Delta z/2$; that is, $\Delta r_{a} = \Delta z$, since our atoms only have a vertical separation. In a uniform external magnetic field, $\mathbf{B_{\rm{ext}}}$, these cylinders will produce an induced magnetic field. The gradients in the induced field will couple to the atom interferometer through the quadratic Zeeman effect to produce an interaction potential that is sensitive to the pendulum's rotation angle and the atom's location.

The magnetic field from a finite length cylinder with uniform magnetization has been calculated in Ref. \cite{CylinderMagnetization}. The total magnetic field is a superposition of the induced field from each cylinder and the external magnetic field. The resulting phase accumulation rate for a range of atom locations and oscillator rotation angles is displayed in Figures \ref{fig:phidot_yz} and \ref{fig:phidot_thetaz}. With the atoms centered at $y=z=0$, the interferometer phase is calculated for the oscillator in classical motion: $\theta(t) = \theta_{\text{max}} \sin{(\omega t)}$. The result is plotted in Fig. \ref{fig:simulatedphase} along with the rotation angle of the oscillator. The interferometer phase, and therefore the probability to find the atom in each lattice site, is clearly correlated with the oscillators position. If the oscillator is quantum mechanical and its position and momentum become sufficiently correlated with the atoms, then there is the possibility of observing the system in a non-classical state.

\subsubsection{Interaction with a quantum harmonic oscillator}

The combined oscillator-atom Hamiltonian has the following form \cite{PRXQuantum.2.030330}:
\begin{equation}
    H =  \omega \hat c^\dagger \hat c + g(\hat c + \hat c^\dagger) \hat S_z
    \label{hamiltonian}
\end{equation}
Here, $\hat c^\dagger$ and $\hat c$ are the oscillator creation and annihilation operators, while $\hat S_{z}$ is a pseudo-spin operator describing the atom's location. The second term is the interaction potential in Eq. \ref{eq:interaction_potential} with quantized oscillator and atom coordinates. $g$ is defined as follows.
\begin{equation}
    g = \sqrt{\frac{1}{2~n_{\rm{cyl}} ~ \pi R^{2} L \rho_{\rm{cyl}}~\omega}} \Delta z \phi_0 \frac{1}{R_0-R}\dfrac{d^2B^2( 0, 0)}{dz ~d\theta}
    \label{eq:g_def}
\end{equation}
Here, $n_{\rm{cyl}}$ is the number of cylinders. The factor of $1/(R_0-R)$ simply arises from taking the derivative in the direction of $\theta$. For oscillator mass and frequency $m$ and $\omega$, the harmonic oscillator position operator is  $\hat r_o = (R_{0}-R)\hat \theta = \sqrt{\frac{1}{2 m \omega}}(\hat c + \hat c ^\dagger)$. Since the atom is confined to two sites in an optical lattice separated by $\Delta z$, the atom's position operator can be expressed as a spin operator in a pseudo-spin-1/2 Hilbert space.
\begin{equation}
    \hat r_a = \Delta z/2 \ket{0}\bra{0} - \Delta z/2 \ket{1}\bra{1} = \Delta z \hat S_z
\end{equation}
For the more realistic case of an atom interferometer with $N$ atoms, the interaction potential can be linearized in each of their coordinates to produce an interaction linear in the collective $N/2$ pseudo-spin operator $\hat{J}_z$ \cite{PRXQuantum.2.030330}.
\begin{equation}
    U_{\text{multi-atom}} = g(\hat c + \hat c ^\dagger) \hat{J}_z
\end{equation}
The system Hamiltonian, Eq. \ref{hamiltonian}, is capable of generating entanglement between the atoms and oscillator. For the oscillator initialized in its ground state and the atom in a symmetric spatial superposition, the system evolves to an entangled state where the displacement of the oscillator is correlated with the atom's location \cite{PRXQuantum.2.030330}, 
\begin{equation}
    \dfrac{\ket{0}_a \ket{\delta}_o + \ket{1}_a\ket{-\delta}_o}{\sqrt{2}}
    \label{after time evolution}
\end{equation}
$|\delta|$ is the coherent state amplitude proportional to $\lambda = g/\omega$. For an experiment to resolve the correlations between the oscillator and the atom, $\lambda$ must be sufficiently large.

We can readily calculate $\lambda$ with Eq. \ref{eq:g_def} for our proposed setup. With the parameter values in Table \ref{tab:PhysLimitsParameters}, we find the following value for $\lambda$ (for $n_{\rm{cyl}}$ = 4) to first order in $\chi_m$:
\begin{equation}
\begin{split}
\ \lambda = -4.0 
\times 10^{-16} \times \left(\frac{B_{\text{ext}}}{5 ~\text{mT}} \right)^2  \left( \frac{ |{\chi_m}|  }{6.8 \times 10^{-5} } \right) \\  \times \left(\frac{19.3 ~\rm{g/cm^{3}}}{\rho_{\rm{cyl}}}\right)^\frac{1}{2} \left( \frac{1}{\alpha} \right)^\frac{7}{2}
\end{split}
\label{eq:LambdaValue}
\end{equation}
The susceptibility of $6.8 \times 10^{-5}$ and density of $19.3~\rm{g/cm^{3}}$ correspond to those of tungsten \cite{Tungsten,TungstenSusc}.
$\alpha$ parametrizes the length scale of the system as $L \rightarrow \alpha L$; that is, $\alpha$ is the scaling factor for all the \textit{length} parameters in Table \ref{tab:PhysLimitsParameters}, with the exception of $\Delta z$ (i.e the parameters $R_{0},l,R,$ and $s$). This scaling implies that $g$ is increased by decreasing the system size. In the parameter space we are at, recalling that $\mathbf{B} = \mathbf{B_{\rm{ext}}}+\mathbf{B_{\rm{cyl}}}$ (with $\mathbf{B_{\rm{cyl}}}$ being the induced magnetic field of the cylinders), the term  $dB^{2}/dz$ of Eq. \ref{eq:g_def} can be approximated to be $2~\mathbf{B}_{\rm{ext}}.(\partial \mathbf{B}_{\rm{cyl}}/\partial z)$. This is because $|\mathbf{B}_{\rm{ext}}|^{2}$ has no gradient and $\partial|\mathbf{B}_{\rm{cyl}}|^{2}/\partial z$ is negligible compared to the cross-term. This allows the scalings of $|\chi_{m}|$ and $\alpha$ to be expressed as ``single" factors. As a side note, the value of $-4.0 
\times 10^{-16}$ was obtained using the expressions in Ref. \cite{CylinderMagnetization}.

The value of the coupling in Eq. \ref{eq:g_def} is extremely small. As will be seen in the succeeding sections, entanglement witness violation requires couplings of around $\lambda \sim 10^{-4.5}$, which is orders of magnitude larger than that which can be obtained from the parameters in Table \ref{tab:PhysLimitsParameters}. There are several ways through which we can increase $\lambda$. One of the primary changes is to switch to a superconducting mass. Though one might think that this will re-introduce the contribution from $\partial|\mathbf{B}_{\rm{cyl}}|^{2}/\partial z$, it is still $\times 10^{2}$ smaller than the cross-term (at least for the physical parameters we use). Therefore, we can still use Eq. \ref{eq:g_def} to inform us of how the values change with rescaling the parameters; having said that, the upcoming calculation was done exactly using the expressions in Ref. \cite{CylinderMagnetization}. We now have $\chi = -1$, which will increase the coupling by a factor of $10^{5}$; increasing the field to 50 mT will contribute a factor of $10^{2}$ (note that at this field, we are still well before the Paschen-Back regime for cesium, as seen in Ref. \cite{SteckCesium}); the system size can be reduced by a factor of 2 (taking $\alpha = 0.5$); a cloud of $N$ atoms leads to a $\sqrt{N}$ increase - we set $N = 10^6$. If we take the material to be lead, then the density will be $\rho_{\rm{cyl}}=11.34$ g/cm$^{3}$ \cite{Lead}. Together, these factors increase the coupling to $\lambda_{N} = 8.8 \times 10^{-6}$. The main concern would be the experimental challenges for some of these changes; in particular, scaling the atom to pendulum distance from 8 mm to 4 mm would be pushing the limits of existing systems, since there is a requirement that the atom trap laser not interact with the pendulum. 

We can also gain improvements in coupling by having the cesium atoms be magnetically sensitive states; for instance, we can place one atom in the $m_{F}=1$ state and the other in the $m_{F}=-1$. In such a scenario, the coupling would primarily be due to the first order Zeeman shift, which is linear in $B$  \cite{SteckCesium}. The energy difference in the atoms now arises due to the difference in their internal states, in addition to the magnetic field differences between their two locations (that is, a coupling will exist even if $\Delta z = 0$). In particular, the term of the Zeeman energy shift which contributes to the coupling is \cite{SteckCesium}
\begin{equation}
   \ \omega_{\rm{1st,~zeeman}} = \frac{(g_{J}-g_{I})~\mu_{B}}{(2 I + 1) \hbar}m_{F}B 
\end{equation}
where $I$ is the nuclear spin, which is 7/2 for Cs; we take the difference of this quantity between the two atoms, and then the gradient in the direction of oscillator motion to obtain the coupling. However, having atoms be located symmetrically in the top and bottom halves of the cylinder is detrimental to the coupling, because now it arises solely from $dB/d\theta$.
Thus, we shift the positions of both atoms such that they are both in $z>0$ plane. If we place the top atom to be at \textit{around} $L/2$, we obtain the coupling to be 
\begin{equation}
\begin{split}
\lambda = -4.4 \times 10^{-12} \times \left(\frac{B_{\text{ext}}}{5 ~\text{mT}} \right)  \left( \frac{ |{\chi_m}|}{6.8 \times 10^{-5}} \right) \\ \left(\frac{19.3 ~\rm{g/cm^{3}}}{\rho_{\rm{cyl}}}\right)^\frac{1}{2} \left( \frac{1}{\alpha} \right)^\frac{5}{2};
\end{split}
\label{eq:LambdaValueFirstZeeman}
\end{equation}
we can see that it is now four orders of magnitude higher. Note that the $\alpha$ scaling here also incorporates changing the position of the top atom location by a factor of $\alpha$; the separation between the two atom locations is still kept constant, so the position of the second location changes too. For 50 mT, $\chi = -1$, $\rho_{\rm{cyl}}=11.34$ g/cm$^{3}$ \cite{Lead}, $N=10^{6}$, and $\alpha = 0.5$, we will get $\lambda_{N} = 4.8 \times 10^{-3}$, which is a significant improvement from having purely $m_{F}=0$ states. Note that scaling $\alpha$ also decreases the atom-oscillator separation from 8 mm to 4 mm. 

If we re-calculate the superconducting case for the simplified setup in the main paper, with changes such that $n_{\rm{cyl}}=2$, the atom-oscillator separation is 8 mm (which is the more feasible situation), and the atoms are symmetrically located at $\pm \Delta z/2$, we get $\lambda_{N} = 6.3 \times 10^{-4}$ (as quoted in the main paper). Potential issues of having the atoms be in magnetically sensitive states is that it can introduce sensitivities to stray magnetic fields.


\bibliography{apssamp}

\end{document}